\documentclass[aps,preprint]{revtex4}
\usepackage{ragged2e}
\usepackage{amsmath}
\usepackage{amssymb}
\usepackage{mathrsfs}
\usepackage{bm}
\usepackage{ulem}
\usepackage{xcolor}
\usepackage{epstopdf}
\usepackage{graphicx}
\usepackage{subfig}
\usepackage{caption}
\usepackage{grffile}
\DeclareCaptionJustification{justified}{\justifying}
\usepackage[colorlinks=true, allcolors=blue]{hyperref}
\usepackage[titletoc]{appendix}

\begin{document}
\preprint{CTP-SCU/2022005}	
\title{Microstructure of Charged AdS Black Hole with Minimal Length Effects}

\author{Ningchen Bai}
\email{bainingchen@stu.scu.edu.cn}
\author{Aoyun He}
\email{heaoyun@stu.scu.edu.cn}
\author{Jun Tao}
\email{taojun@scu.edu.cn}

\affiliation{Center for Theoretical Physics, College of Physics, Sichuan University, Chengdu, 610065, China}

\begin{abstract}
In this work, the microstructure of charged AdS black holes under minimal length effects is investigated. We study the thermodynamics of black holes in the extended phase space, where the cosmological constant is regarded as the thermodynamic pressure. The modified Hawking temperature and phase transition are obtained based on the generalized uncertainty principle (GUP). Then, using thermodynamic geometry, we show that the microstructure of black holes can be determined by the ratio of GUP parameter to charge. For a small ratio, the black hole exhibits the typical RN-AdS microstructure with van der Waals phase transition and repulsive/attractive interactions. As the ratio increases, the reentrant phase transition takes place, and both the repulsion-attraction coexisted black hole and the attraction dominated black hole can be found in this case. For a large ratio, the black hole behaves like a Schwarzchild-AdS black hole in which neither phase transition nor repulsive interaction exists. These results suggest that the GUP effect will reduce the repulsive interaction presented by the charged AdS black hole, which can also be qualitatively understood from the perspective of black hole molecules.
\end{abstract}

\maketitle
\newpage 


\section{Introduction}\label{section1}
With the well-defined temperature of a black hole \cite{Bekenstein:1973ur,Bekenstein:1974ax,Hawking:1975vcx,Hawking:1976de, Bardeen:1973gs,Gibbons:1976ue}, it is reasonable to explore the associated microscopic structure. In spite of the great progress made by string theory \cite{Strominger:1996sh,Maldacena:1996gb,Callan:1996dv,Horowitz:1996fn}, fuzzy ball model \cite{Lunin:2001jy,Lunin:2002qf}, and loop quantum gravity \cite{Rovelli:1996dv}, the microstructure of black holes remains a tricky problem. Under such circumstances, we hope to draw inspiration from study on the thermodynamics of black holes which might reveal more microscopic characteristics.

In recent years, the study on the thermodynamics of AdS black holes in extended phase space has received great attention, where the cosmological constant is regarded as the thermodynamic pressure \cite{Kastor:2009wy,Dolan:2010ha,Dolan:2011xt,Cvetic:2010jb}. Certain AdS black holes exhibit a phase transition that is analogous to that of a van der Waals (VdW) fluid \cite{Kubiznak:2012wp,Gunasekaran:2012dq,Kubiznak:2016qmn, He:2016fiz}. In general, phase transitions are intimately tied to the microscopic characteristics of the system, which suggests that AdS black holes have fluid-like microstructures. Numerous phase transition behaviors displayed by the AdS black holes provide additional support for this perspective, such as triple point \cite{Altamirano:2013uqa,Wei:2014hba,Frassino:2014pha}, reentrant phase transition \cite{Altamirano:2013ane} and superfluid black hole phase \cite{Hennigar:2016xwd}. 

For a standard fluid system, microscopic characteristics can be directly identified from its molecular constituents, followed by macroscopic thermodynamics using statistical mechanics. Nevertheless, the microscopic structure of black holes remains unclear and thus the thermodynamic geometry plays an important role. Inspired by Ruppiener \cite{Ruppeiner:1995zz} and the pioneering work of Weinhold \cite{weinhold1975metric} in standard thermodynamics, it has been confirmed that the scalar curvature associated with such thermodynamic geometry is connected to the microscopic interactions, with positive/negative scalar curvature implying repulsive/attractive interactions \cite{ruppeiner1979thermodynamics,ruppeiner1981application,janyszek1989riemannian,janyszek1990riemannian,oshima1999riemann,mirza2009nonperturbative,may2013thermodynamic}. It is straightforward to apply Ruppeiner geometry to black hole systems since entropy is treated as thermodynamic potential. In Refs. \cite{Wei:2019uqg,Wei:2019yvs,Dehyadegari:2020ebz}, the Ruppeiner geometry of RN-AdS black holes has been explored and intriguing results on its microscopic description are reported. Apart from a dominant attractive interaction for most parameter ranges, a repulsive interaction was discovered for small black hole (SBH) with high temperature. Along the coexistence curve, the large black hole (LBH) branch always exhibits dominant attractive interaction, while the SBH branch exhibits a transition between dominant repulsive and attractive interactions at low temperature, indicating a dramatic change in the black hole's microscopic interaction during the phase transition. This result is considerably distinct from that of van der Waals systems, which only present attractive interaction between their microscopic molecules \cite{Wei:2019yvs}. Such a method combining phase transition and Ruppeiner geometry has also been used to investigate the microstructures of several other black hole systems \cite{Ghosh:2019pwy,Wei:2019ctz,Xu:2020gud,Ghosh:2020kba,NaveenaKumara:2020jmk,Wei:2020poh,NaveenaKumara:2020biu,Yerra:2020oph,Wei:2021lmo,Gogoi:2021syo,Yerra:2021hnh,Yerra:2020tzg,Dehyadegari:2021ieh}.

On the other hand, it has been observed that the generalized uncertainty principle (GUP) \cite{Kempf:1994su,Kempf:1996nk} has a crucial influence on the thermodynamics of black holes. One important effect is that the black hole mass cannot be lower than the order of Planck mass, implying the existence of a black hole remnant \cite{Custodio:2003jp,Nouicer:2007jg,Nouicer:2007pu,Kim:2007hf}. In addition, as anticipated by string theory and loop quantum gravity, the GUP modifies the usual Hawking temperature and Bekenstein-Hawking area law of entropy \cite{Adler:2001vs,Medved:2004yu,Nozari:2006ka,Nozari:2006vn,Bina:2010ir,Chen:2013ssa,Wang:2015bwa,Guo:2015ldd,Iorio:2019wtn}. The significant influence of GUP on the black hole phase transitions has also been confirmed \cite{Myung:2006qr,Sabri:2012fi,Ma:2018svr,Li:2018vgt,Fu:2021zrd}. Considering the relation between thermodynamics and microstructures, we hope to find out what the underlying microstructure in a black hole with GUP effect is. In this article, the microstructure of the GUP corrected charged AdS black hole will be probed in terms of Ruppeiner geometry and phase transition behavior in the extended phase space.

This paper is structured as follows. Sec. \ref{section2} reviews the GUP-derivation thermodynamics of a charged AdS black hole. Then, Sec. \ref{section3} studies the black hole's phase transition behavior. The cases of VdW like phase transition and reentrant phase transition are discussed separately. Sec. \ref{section4} is devoted to exploring microstructures of the black hole by use of Ruppeiner geometry. Finally, Sec. \ref{section5} contains a summary of our findings.

\section{Thermodynamics for GUP-corrected charged AdS black holes}\label{section2}	
In four-dimensional spacetime, the metric of charged AdS black holes is
\begin{equation}
\label{eq:1}
ds^2=-f(r)dt^2+\frac{1}{f(r)}dr^2+r^2d\Omega^2_{2},
\end{equation}
with the metric function given by
\begin{equation}
\label{eq:2}
\begin{aligned}
f(r)=1-\frac{2M}{r}+\frac{Q^2}{r^2}-\frac{\Lambda r^2}{3}.
\end{aligned}
\end{equation}
The parameter $\Lambda$ is the cosmological constant, and $M$, $Q$ represent the ADM mass and electric charge of the black hole respectively.

The Hawking temperature and entropy of the black hole in the semiclassical regime are obtained as
\begin{equation}
\label{eq:3}
T=\frac{\hbar \kappa}{2\pi},\quad S=\frac{A}{4\hbar},
\end{equation}
where $A=4\pi r^2_{h}$ represents the area of the horizon, and $\kappa$ is the surface gravity of the black hole given by
\begin{equation}
\label{eq:4}
\kappa=\frac{f^{\prime}\left(r_{h}\right)}{2}=\frac{r^2_{h}+8\pi Pr^4_{h}-Q^2}{2r^3_{h}}.
\end{equation}

In the extended phase space, we interpret the cosmological constant $\Lambda$ as the thermodynamic pressure \cite{Kastor:2009wy,Dolan:2010ha,Dolan:2011xt,Cvetic:2010jb}
\begin{equation}
\label{eq:5}
P=-\frac{\Lambda}{8\pi},
\end{equation}
and the first law of black hole thermodynamics reads 
\begin{equation}
\label{eq:6}
dM=TdS+VdP+\Phi dQ,
\end{equation}
where $\Phi=Q/r_{h}$ is the electric potential on the horizon $r_{h}$ measured at infinity, and $V=4\pi r^3_{h}/3$ is the thermodynamic volume of black hole.

In more general cases, the entropy of a black hole should be a function of $A$, i.e. $S=S(A)$ \cite{Bekenstein:1973ur}. As a result, the temperature can be re-calculated as \cite{Xiang:2009yq}
\begin{equation}
\label{eq:7}
T=\left(\frac{\partial M}{\partial S}\right)_{P, Q}=\frac{dA}{dS}\times 	\left(\frac{\partial M}{\partial A}\right)_{P, Q}=\frac{dA}{dS}\times \frac{\kappa}{8\pi}.
\end{equation}

Considering the following GUP effect 
\begin{equation}
\label{eq:8}
\Delta x\geq\frac{\hbar}{\Delta p}+\frac{\alpha^2}{\hbar}\Delta p,
\end{equation}
and the process of a particle being absorbed by the black hole, one can show that \cite{Xiang:2009yq}
\begin{equation}
\label{eq:9}
\frac{d A}{d S}\simeq \frac{8 \hbar}{\alpha^{2}}\left(r_{h}^{2}-r_{h} \sqrt{r_{h}^{2}-\alpha^{2}}\right).
\end{equation}

Substituting Eqs. (\ref{eq:9}) and (\ref{eq:4}) into Eq. (\ref{eq:7}), we have
\begin{equation}
\label{eq:10}
T=\frac{\hbar\left(r_{h}-\sqrt{r_{h}^{2}-\alpha^{2}}\right)\left( r_{h}^{2}+8 \pi Pr^{4}_{h}-Q^{2}\right)}{2 \pi \alpha^{2} r_{h}^{2}}.
\end{equation}

The GUP-corrected black hole entropy can also be computed as
\begin{equation}
\label{eq:11}
\begin{aligned}
S &=\int \frac{d S}{d A} d A + S_{0} \\ &=\frac{\pi}{2 \hbar}\left[r_{h}^{2}+r_{h} \sqrt{r_{h}^{2}-\alpha^{2}}-\alpha^{2} \ln \left(\frac{\sqrt{r_{h}^{2}-\alpha^{2}}+r_{h}}{\alpha}\right)\right],
\end{aligned}
\end{equation}
where $S_{0}$ is an integral constant. Following Ref. \cite{Ma:2018svr}, we take $S_{0}=\alpha^{2} \ln \alpha$ to obtain a dimensionless logarithmic term.

Eq. (\ref{eq:10}) gives a GUP-corrected state equation as follows
\begin{equation}
\label{eq:12}
P=\frac{r_{h}+\sqrt{r_{h}^{2}-\alpha^{2}}}{4r^2_{h}}T-\frac{1}{8\pi r^2_{h}}+\frac{Q^2}{8\pi r^4_{h}},
\end{equation}
where we have set $\hbar=1$. It can be expanded in a series of the inverse of horizon radius $r_{h}$ 
\begin{equation}
\label{eq:13}
P=\frac{T}{2 r_{h}}-\frac{1}{2\pi (2r_{h})^{2}}-\frac{\alpha ^2 T}{(2r_{h})^{3}}+\frac{2 Q^{2}}{\pi (2 r_{h})^{4}}-\frac{\alpha ^4 T}{(2r_{h})^{5}}+\mathcal{O}\left((2 r_{h})^{-6}\right).
\end{equation}
Comparing this state equation with the virial expansion of VdW equation, the specific volume can be identified as $v=2 r_{h}$, and the state equation becomes
\begin{equation}
\label{eq:14}
P=\frac{T}{v}\left[\frac{1}{2}+\frac{\sqrt{v^{2}-4 \alpha^{2}}}{2 v}\right]-\frac{1}{2 \pi v^{2}}+\frac{2 Q^{2}}{\pi v^{4}}.
\end{equation}

Setting $\alpha=0$, it reduces to the semi-classical state equation of charged AdS black holes \cite{Kubiznak:2012wp}
\begin{equation}
\label{eq:15}
P=\frac{T}{v}-\frac{1}{2 \pi v^{2}}+\frac{2 Q^{2}}{\pi v^{4}}.
\end{equation}

If we express the state equation as the following form
\begin{equation}
\label{eq:16}
P=a(V)T+b(V,Q),
\end{equation}
where $V=\pi v^{3}/6$, we can find that the GUP will not change the $b(V, Q)$ term. Furthermore, from Eq. (\ref{eq:10}), the GUP-corrected $T$ gives a mandatory requirement $r_{h}\geqslant \alpha$, i.e. $v\geqslant 2\alpha$. These features have great influence on the microstructure of charged AdS black hole as we will see soon. On the other hand, we emphasize that the GUP only gives correction to the temperature and the entropy by constraining the minimal length, not the electric charge and the electric potential, and thus the first law of black hole thermodynamics remains valid.

\section{Phase transition of GUP-corrected charged AdS black holes}\label{section3}
The critical point of Eq. (\ref{eq:14}) is obtained by employing the equations
\begin{equation}
\label{eq:17}
\frac{\partial P}{\partial v}=0,\quad \frac{\partial^{2} P}{\partial v^{2}}=0.
\end{equation}
Combining them, a constraint equation is given as \cite{Ma:2018svr}
\begin{equation}
\label{eq:18}
\begin{aligned}
\frac{\alpha^{2}}{Q^{2}}=\frac{2\left(4 \beta^{5}-\beta^{4}-5 \beta^{3}+2 \beta^{2}+\beta-1\right)}{-\beta^{2}+\beta-1},
\end{aligned}
\end{equation}
where $\beta=\sqrt{1-4\alpha^{2} / v^{2}}$. The right-hand side of Eq. (\ref{eq:18}) is shown in Fig. \ref{fig:Fig.1}. The value of $\alpha^{2}/Q^{2}$ determines the number of critical points. Based on the discussion in Ref. \cite{Ma:2018svr}, the system can be classified into three different cases according to the value of $\alpha^{2}/Q^{2}$:

\begin{figure}[h]
\centering
\includegraphics[width=0.45\textwidth]{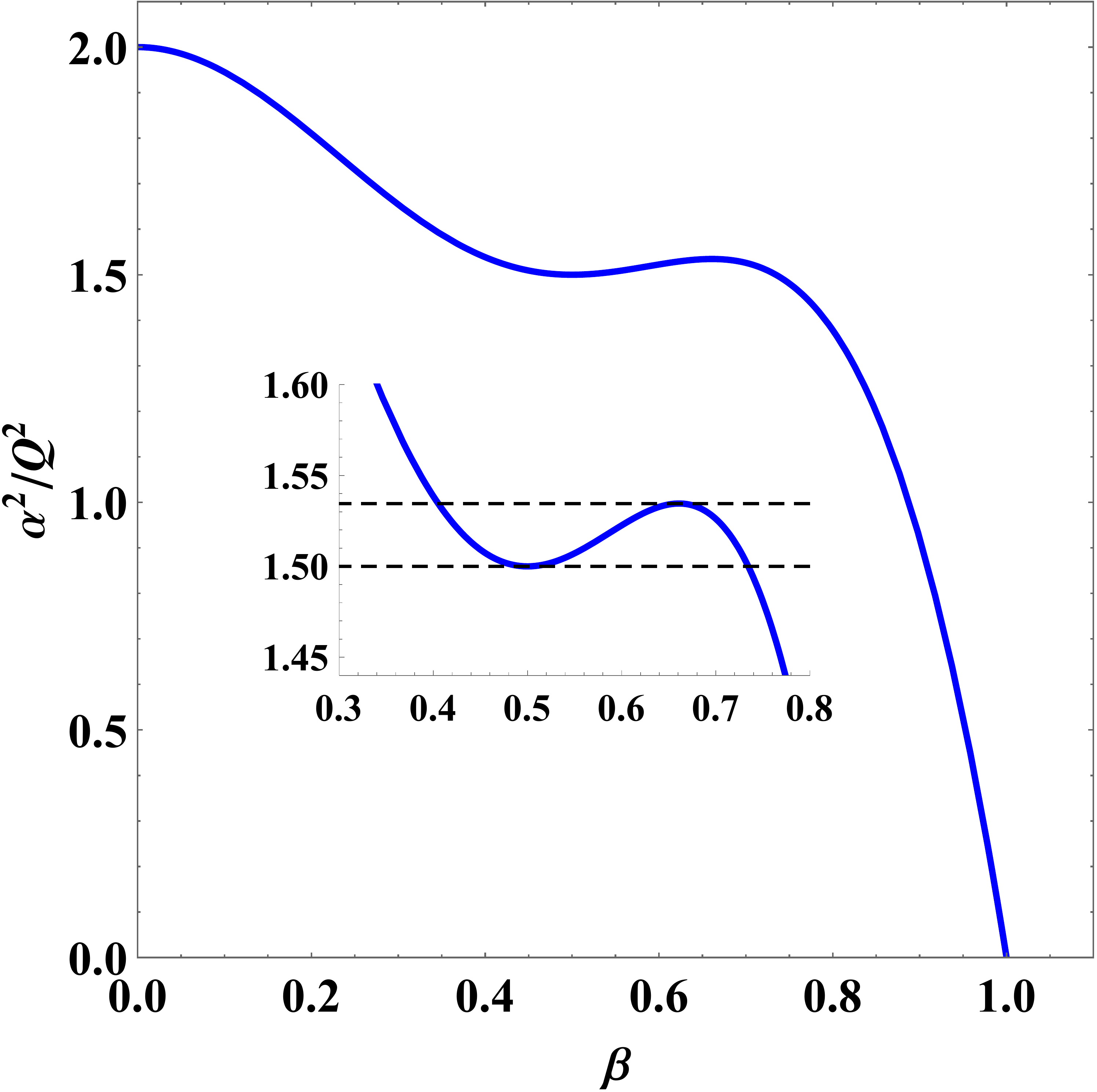}
\caption{\label{fig:Fig.1}Number of critical points for the GUP-corrected charged AdS black hole.}
\end{figure}

\noindent \textbf{VdW-like PT Case:} $\alpha^{2}/Q^{2} \leqslant 1$. The system exhibits a typical VdW-like first-order phase transition, similar to the RN-AdS black hole. Unlike the phase transition between SBH and LBH, this transition occurs between an intermediate black hole (IBH) phase and a LBH phase. There is only one critical point in this case.

\quad

\noindent \textbf{RPT Case:} $1<\alpha^{2}/Q^{2}<1.535$. A zeroth-order phase transition and a VdW-like first-order phase transition are present in the system, between the IBH phase and the LBH phase. These successive phase transitions together are referred to a reentrant phase transition (RPT). Specially, for $1<\alpha^{2}/Q^{2}<1.5$, the system only contains one critical point. When $\alpha^{2}/Q^{2}=1.5$, there are two critical points $c$ and $c_{1}$ in the system. However, only the critical point $c$ with higher pressure has physical significance, which undergoes a second-order phase transition. For $1.5<\alpha^{2}/Q^{2}<1.535$, three critical points $c$, $c_{1}$ and $c_{2}$ occur in the system, although only the critical point $c$ with higher pressure has physical significance.

\quad

\noindent \textbf{No PT Case:} $\alpha^{2}/Q^{2} \geqslant 1.535$. There is no phase transition in the system, as the LBH is globally stable. For $\alpha^{2}/Q^{2}=1.535$, there are two critical points $c_{1}$ and $c_{2}$ in the system. The pressure at critical point $c_{2}$ is lower, and negative. For $1.535<\alpha^{2}/Q^{2} \leqslant 2$, the system contains one critical point with negative pressure. Specially, when $\alpha^{2}/Q^{2} > 2$, there is no critical point.

\begin{figure}
\centering
\subfloat[$\alpha^{2}/Q^{2}= 0$]
{\begin{minipage}[b]{.49\linewidth}
\centering
\includegraphics[height=7cm]{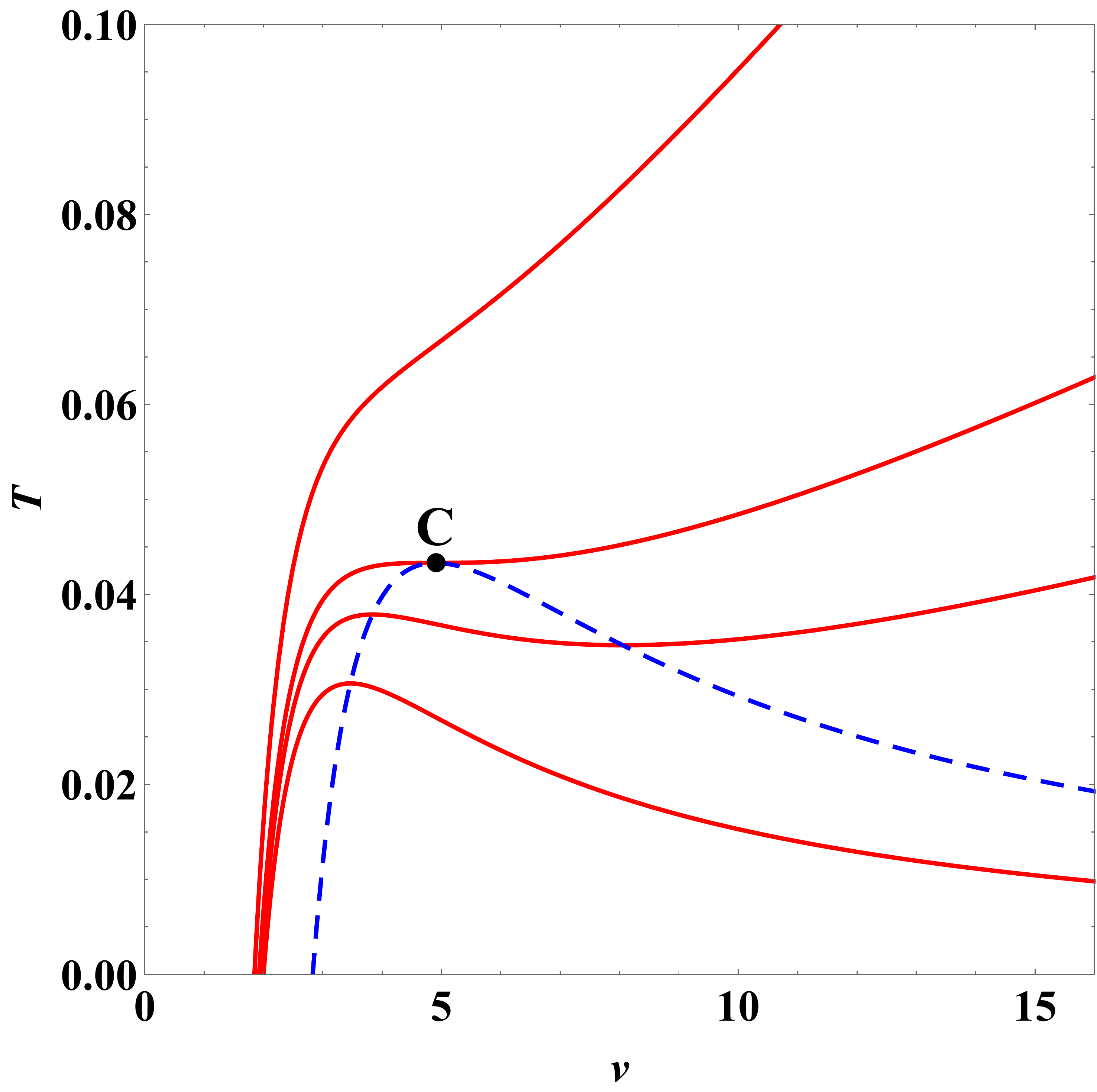}
\label{fig:Fig.2(a)}
\end{minipage}
} 
\subfloat[$\alpha^{2}/Q^{2}=0.95$]
{\begin{minipage}[b]{.49\linewidth}
\centering
\includegraphics[height=7cm]{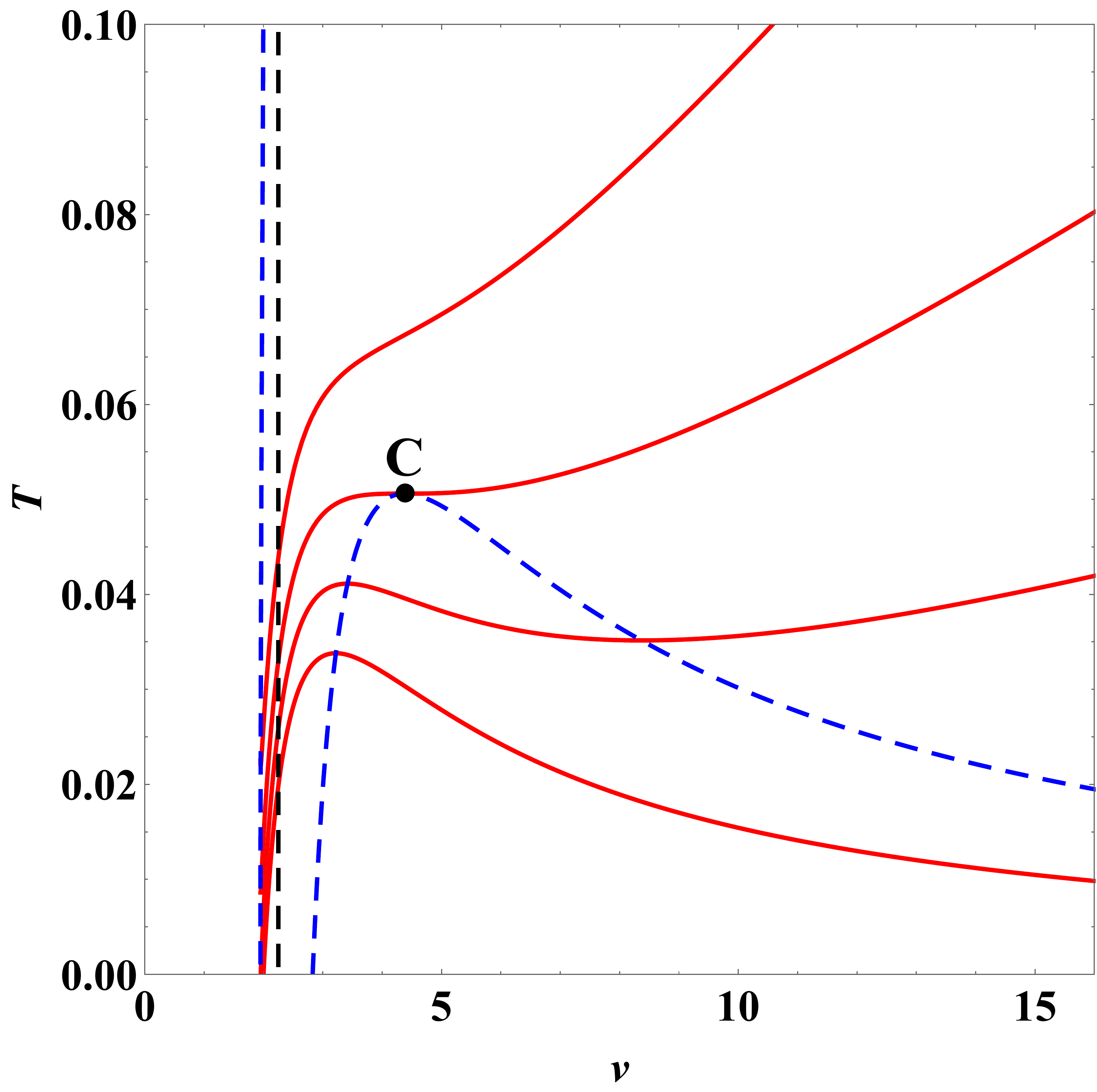}
\label{fig:Fig.2(b)}
\end{minipage}
}
\caption{\label{fig:Fig.2}Isobaric curves of the GUP-corrected charged AdS black hole in VdW-like PT case. (a) Isobaric curves for $\alpha^{2}/Q^{2}= 0$. The pressures are set as $P=0$, $0.002$, $0.0033$ $(P_{c})$ and $0.008$ from bottom to top. (b) Isobaric curves for $\alpha^{2}/Q^{2}= 0.95$. The pressures are set as $P=0$, $0.002$, $0.0044$ $(P_{c})$ and $0.008$ from bottom to top. We have set $Q = 1$.}
\end{figure}

\begin{figure}
\centering
\subfloat[$\alpha^{2}/Q^{2}= 1.4$]
{\begin{minipage}[b]{.33\linewidth}
\centering
\includegraphics[height=5cm]{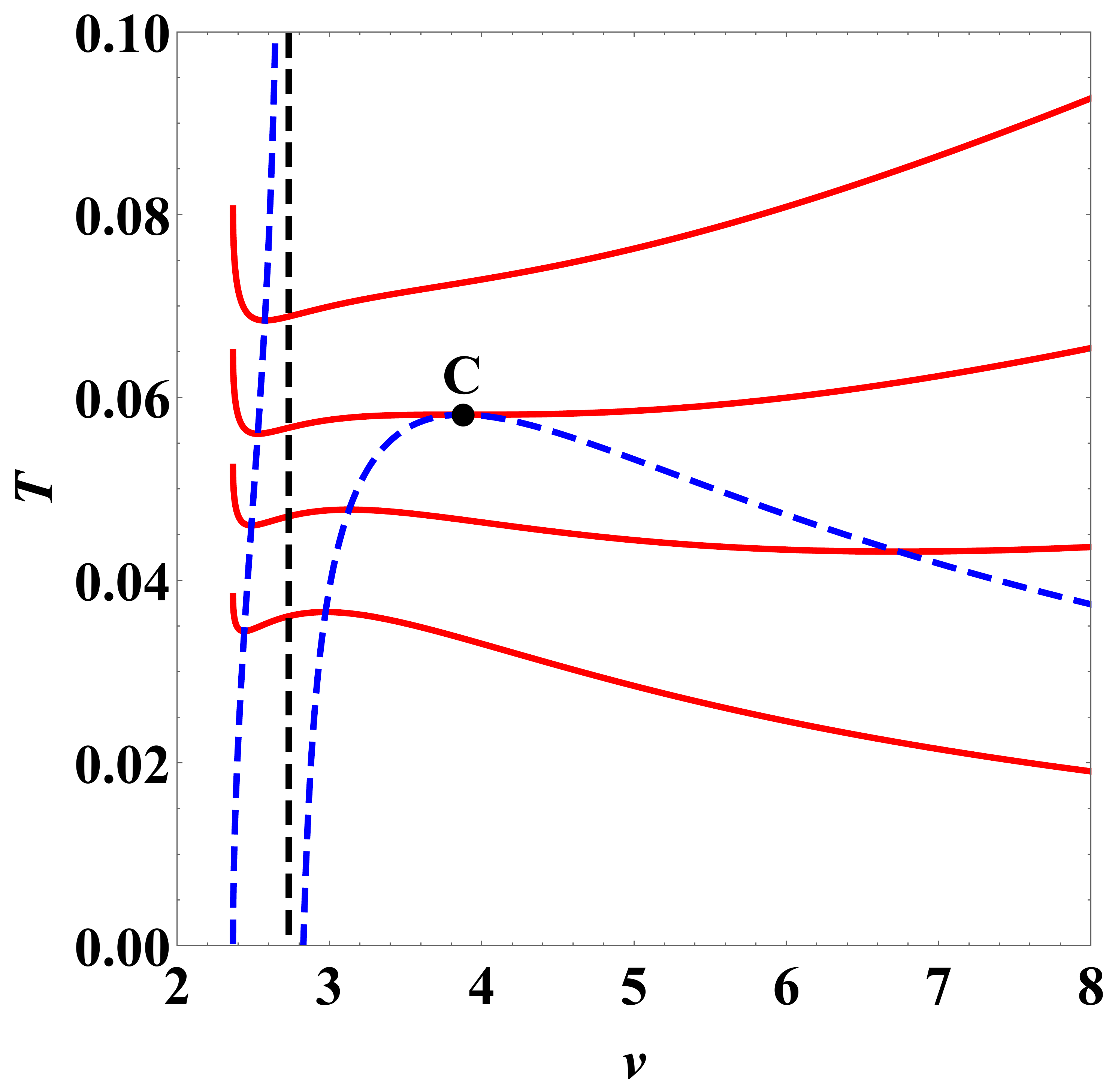}
\label{fig:Fig.3(a)}
\end{minipage}
} 
\subfloat[$\alpha^{2}/Q^{2}=1.5$]
{\begin{minipage}[b]{.33\linewidth}
\centering
\includegraphics[height=5cm]{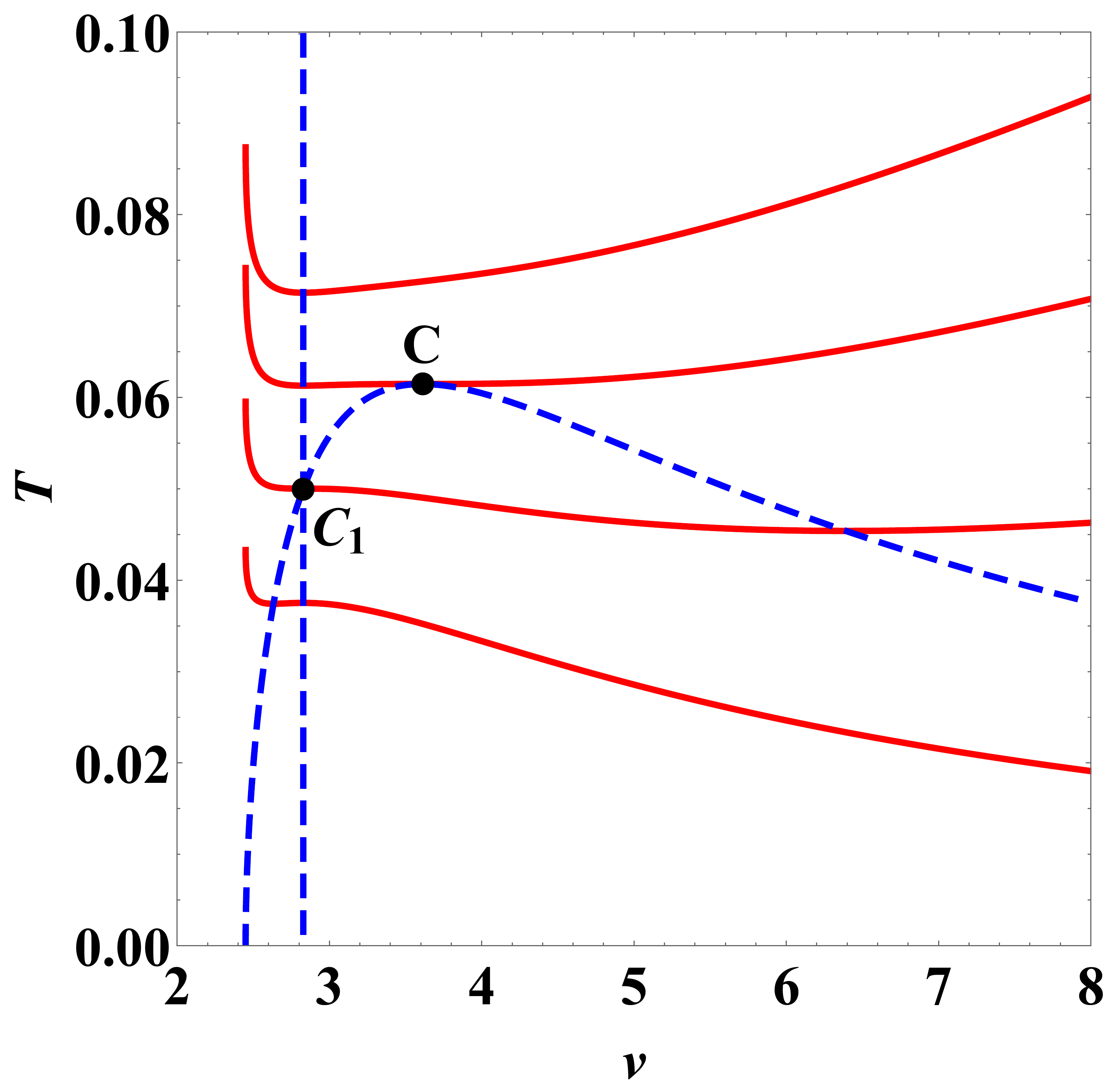}
\label{fig:Fig.3(b)}
\end{minipage}
}
\subfloat[$\alpha^{2}/Q^{2}=1.51$]
{\begin{minipage}[b]{.33\linewidth}
\centering
\includegraphics[height=5cm]{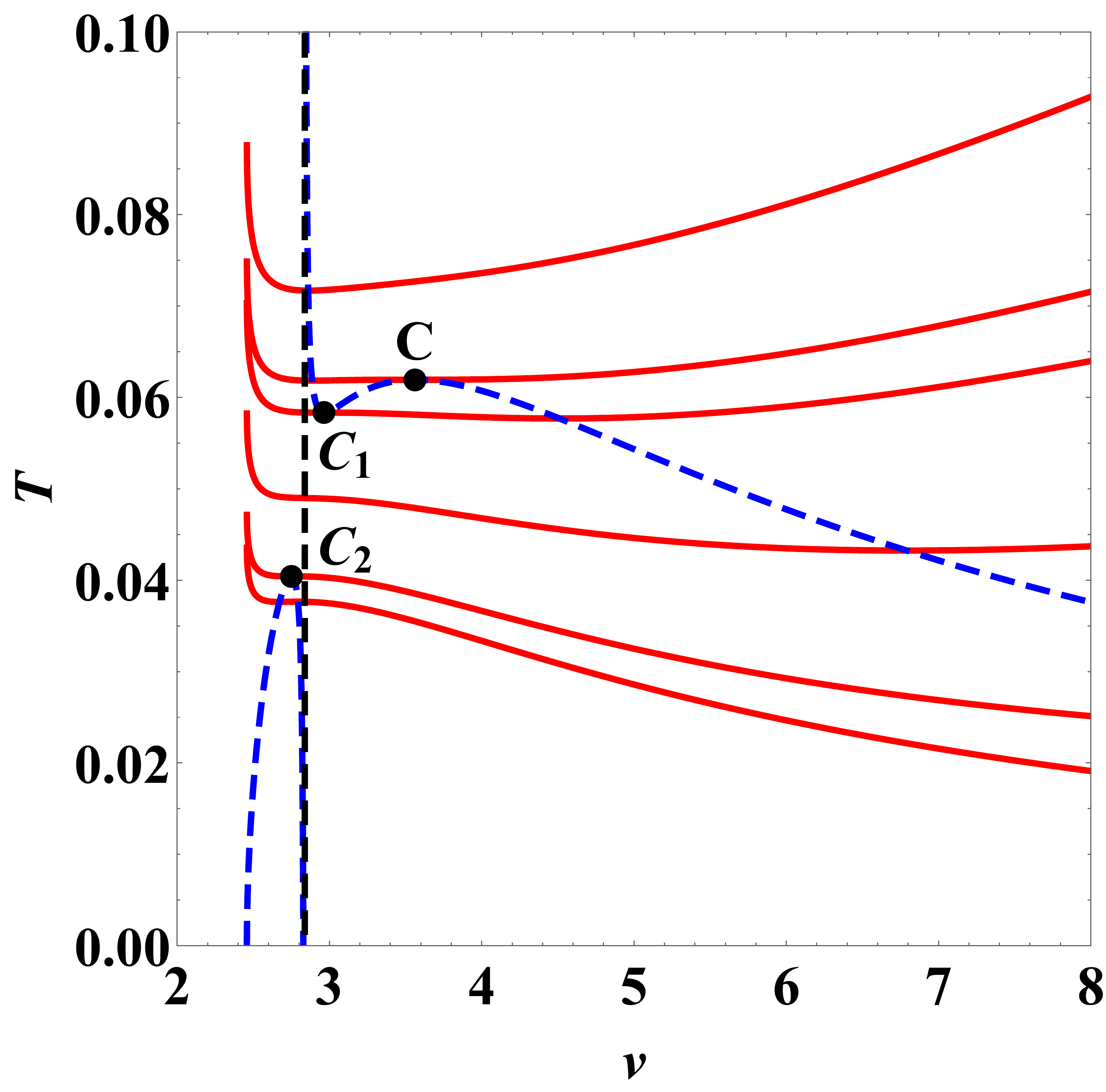}
\label{fig:Fig.3(c)}
\end{minipage}
}
\caption{\label{fig:Fig.3}Isobaric curves of the GUP-corrected charged AdS black hole in RPT case. (a) Isobaric curves for $\alpha^{2}/Q^{2}= 1.4$. The pressures are set as $P=0$, $0.003$, $0.0057$ $(P_{c})$ and $0.009$ from bottom to top. (b) Isobaric curves for $\alpha^{2}/Q^{2}= 1.5$.  The pressures are set as $P=0$, $0.0033$ $(P_{c_{1}})$, $0.0063$ $(P_{c})$ and $0.009$ from bottom to top. (c) Isobaric curves for $\alpha^{2}/Q^{2}= 1.51$. The pressures are set as $P=0$, $0.00073$ $(P_{c_{2}})$, $0.003$, $0.0055$ $(P_{c_{1}})$, $0.0064$ $(P_{c})$ and $0.009$ from bottom to top. We have set $Q = 1$.}
\end{figure}

\begin{figure}
\centering
\subfloat[$\alpha^{2}/Q^{2}= 1.535$]
{\begin{minipage}[b]{.33\linewidth}
\centering
\includegraphics[height=4.95cm]{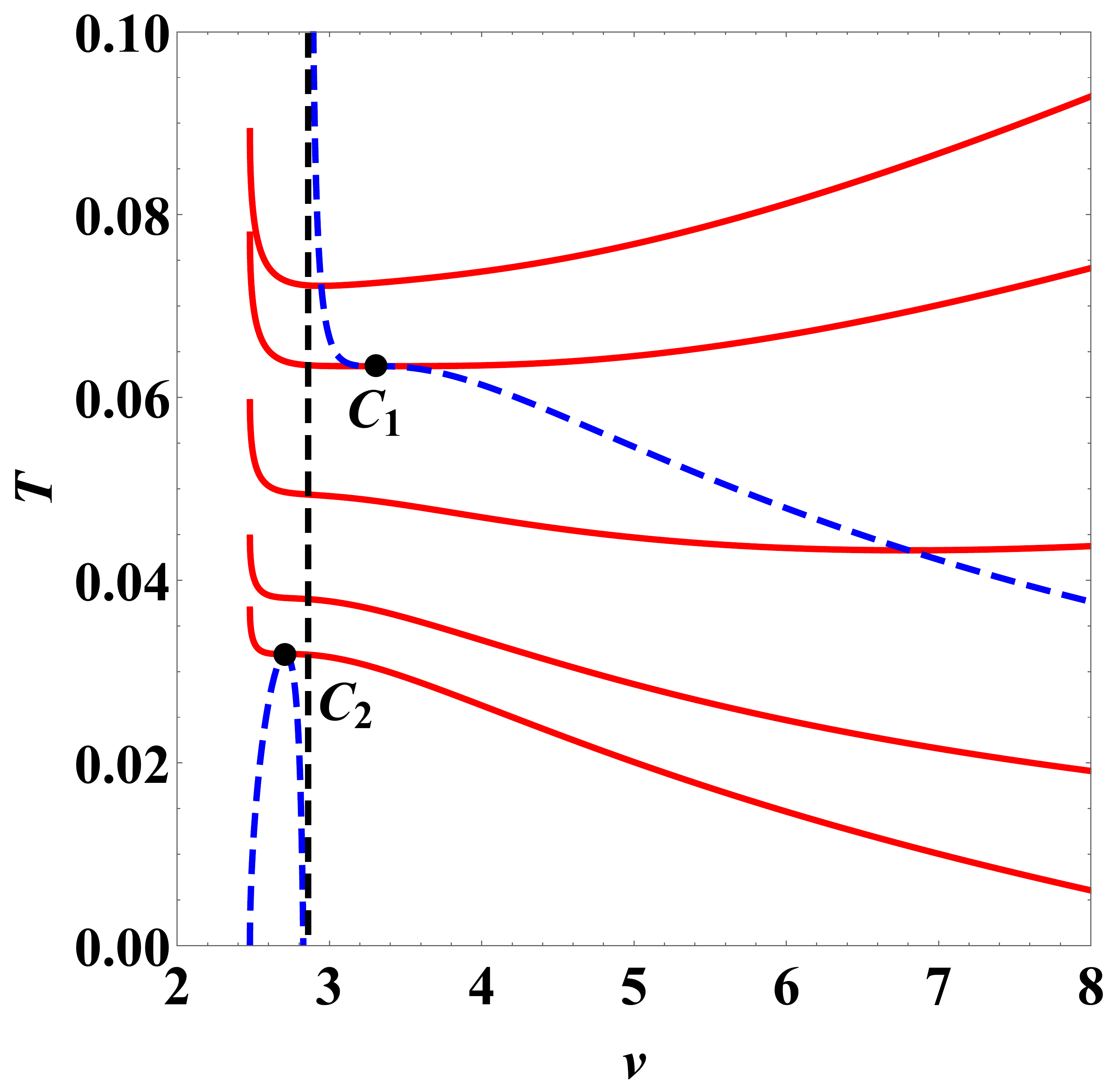}
\label{fig:Fig.4(a)}
\end{minipage}
} 
\subfloat[$\alpha^{2}/Q^{2}=1.6$]
{\begin{minipage}[b]{.33\linewidth}
\centering
\includegraphics[height=4.95cm]{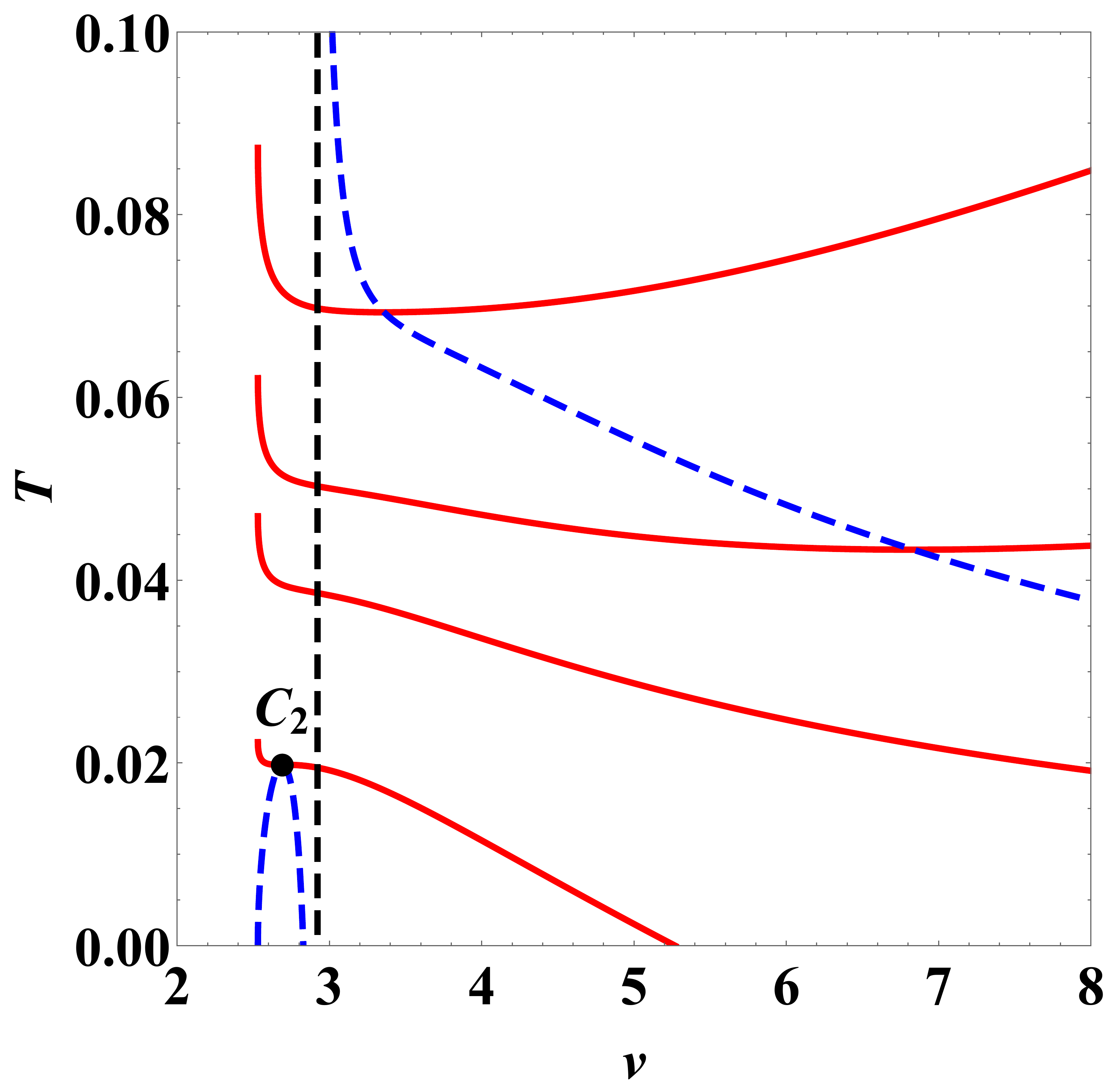}
\label{fig:Fig.4(b)}
\end{minipage}
}
\subfloat[$\alpha^{2}/Q^{2}=2.1$]
{\begin{minipage}[b]{.33\linewidth}
\centering
\includegraphics[height=4.95cm]{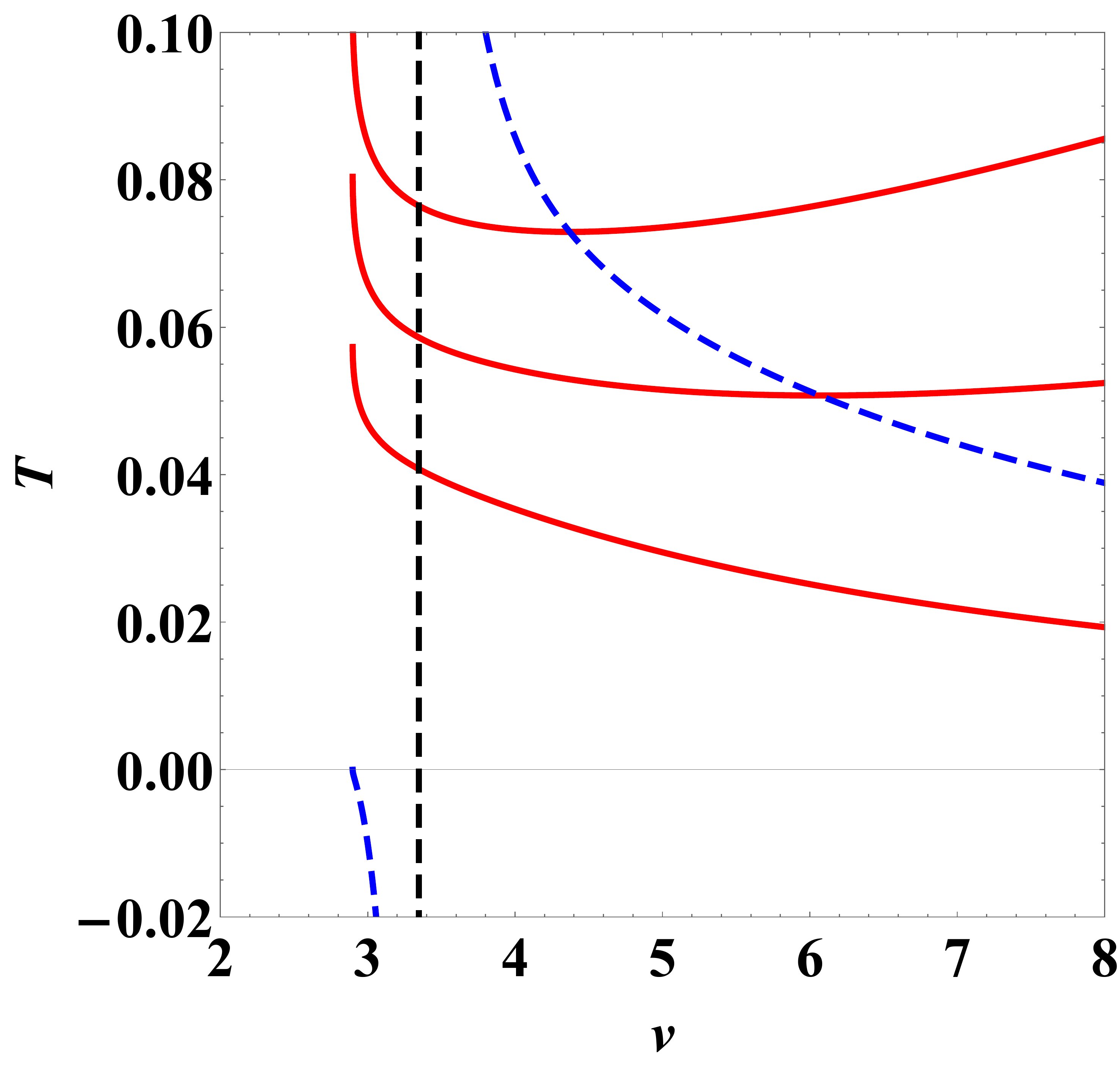}
\label{fig:Fig.4(c)}
\end{minipage}
}
\caption{\label{fig:Fig.4}Isobaric curves of the GUP-corrected charged AdS black hole in no PT case. (a) Isobaric curves for $\alpha^{2}/Q^{2}= 1.535$. The pressures are set as $P=-0.0016$ $(P_{c_{2}})$, $0$, $0.003$, $0.0067$ $(P_{c_{1}})$ and $0.009$ from bottom to top. (b) Isobaric curves for $\alpha^{2}/Q^{2}= 1.6$. The pressures are set as $P=-0.0049$ $(P_{c_{2}})$, $0$, $0.003$ and $0.008$ from bottom to top. (c) Isobaric curves for $\alpha^{2}/Q^{2}= 2.1$. The pressures are set as $P=0$, $0.004$ and $0.008$ from bottom to top. We have set $Q = 1$.}
\end{figure}

For simplicity, we take $Q=1$ hereafter in this
paper. We show the behaviors of the isobaric curves for the VdW-like PT case, RPT case and no PT case in Figs. \ref{fig:Fig.2}--\ref{fig:Fig.4}, respectively. The features discussed above can be found. For comparison, the $\alpha^{2}/Q^{2}= 0$ case, namely the case for RN-AdS black hole, is given in Fig. \ref{fig:Fig.2}(\subref*{fig:Fig.2(a)}). The blue solid lines in these figures represent the spinodal curves specified by
\begin{equation}
\label{eq:19}
\begin{aligned}
\left(\frac{\partial P}{\partial v}\right)_{T}=0, \quad \text { or } \quad\left(\frac{\partial T}{\partial v}\right)_{P}=0.
\end{aligned}
\end{equation}
When $\alpha^{2}/Q^{2} \neq 1.5$, only one spinodal curve is found in $T-v$ diagrams
\begin{equation}
\label{eq:20}
\begin{aligned}
T_{sp}=\frac{2 \left(v^2-8 Q^2\right) \sqrt{v^2-4 \alpha ^2}}{\pi  v^2 \left(v^2-8 \alpha ^2+v \sqrt{v^2-4 \alpha ^2}\right)}.
\end{aligned}
\end{equation}
For $\alpha^{2}/Q^{2} = 1.5$, one can obtain another spinodal curve expressed as
\begin{equation}
\label{eq:21}
\begin{aligned}
v_{sp}=2\sqrt{2}Q.
\end{aligned}
\end{equation}
The numerator of $T_{sp}$ equals zero when $v=2\sqrt{2} Q$ or $2\alpha$, while the denominator reaches zero at $v=4\sqrt{3} \alpha /3$. For this, $T_{sp}$ always equals zero at $v=2\sqrt{2} Q$ or $2\alpha$ and diverges at $v=4\sqrt{3} \alpha /3$ (black dashed lines in Figs. \ref{fig:Fig.2}--\ref{fig:Fig.4}) when $\alpha^{2}/Q^{2} \neq 1.5$. The line $v=2\sqrt{2} Q$ becomes another spinodal curve $v_{sp}$ when $\alpha^{2}/Q^2=1.5$, overlapping with the line $v=4\sqrt{3} \alpha /3$, and it is easy to find that two spinodal curves intersect at $T \equiv \sqrt{2} /(9 \pi Q) \approx 0.05Q$ (see Fig. \ref{fig:Fig.3}(\subref*{fig:Fig.3(b)})). Thus in all cases the line $v=4\sqrt{3} \alpha /3$ separates the spinodal curve $T_{sp}$ into the left and right branches.

These spinodal curves are related to the local thermodynamic stability of a black hole, which can be reflected by the heat capacity at constant pressure, i.e.,
\begin{equation}
\label{eq:22}
\begin{aligned}
C_{P}=T\left(\frac{\partial S}{\partial T}\right)_{P,Q}=T\left(\frac{\partial S/\partial v}{\partial T/\partial v}\right)_{P,Q}.
\end{aligned}
\end{equation}
The black hole branch in the $T-v$ diagram with positive slope is therefore locally stable, while the branch with negative slope is locally unstable. From Eq. (\ref{eq:19}), we conclude that the spinodal curves separate the black hole branches into locally stable and unstable ones. Specially, the locally stable phases are above the right branch of spinodal curve $T_{sp}$, and on the contrary for the left branch, as shown in Figs. \ref{fig:Fig.2}--\ref{fig:Fig.4}. With the increasing of $\alpha$, the locally stable phases will shift to the right along with the spinodal curve.

However, notice that not all of these locally stable branches are globally stable. The globally stable branch is the one with a lower Gibbs free energy, given by
\begin{equation}
\label{eq:23}
\begin{aligned}
G &=M-T S \\
&=\frac{1}{24 v^{2}}\Big[36 Q^{2} v+3 v^{3}-2 P \pi v^{5}\\
& \quad -3\left(-4Q^{2}+v^{2}+2 P \pi v^{4}\right)\left(-v+\sqrt{v^{2}-4\alpha^{2}}\right) \log \left(\frac{v+\sqrt{v^{2}-4\alpha^{2}}}{2\alpha}\right)\Big].
\end{aligned}
\end{equation}
Phases with a higher Gibbs free energy may be meta-stable or unstable, irrespective of whether they have a positive or negative heat capacity. In the followings, the phase transition behavior of the black hole will be studied by the Gibbs free energy. The VdW-like PT case and RPT case are discussed separately. Note that only a stable LBH phase exists in the system when $\alpha^{2}/Q^{2}\geqslant 1.535$ (see Fig. \ref{fig:Fig.4}), and thus no phase transition occurs.

\subsection{Van der Waals-Like Phase Transition}
\begin{figure}[h]
\centering
\subfloat[]
{\begin{minipage}[b]{.33\linewidth}
\centering
\includegraphics[height=5cm]{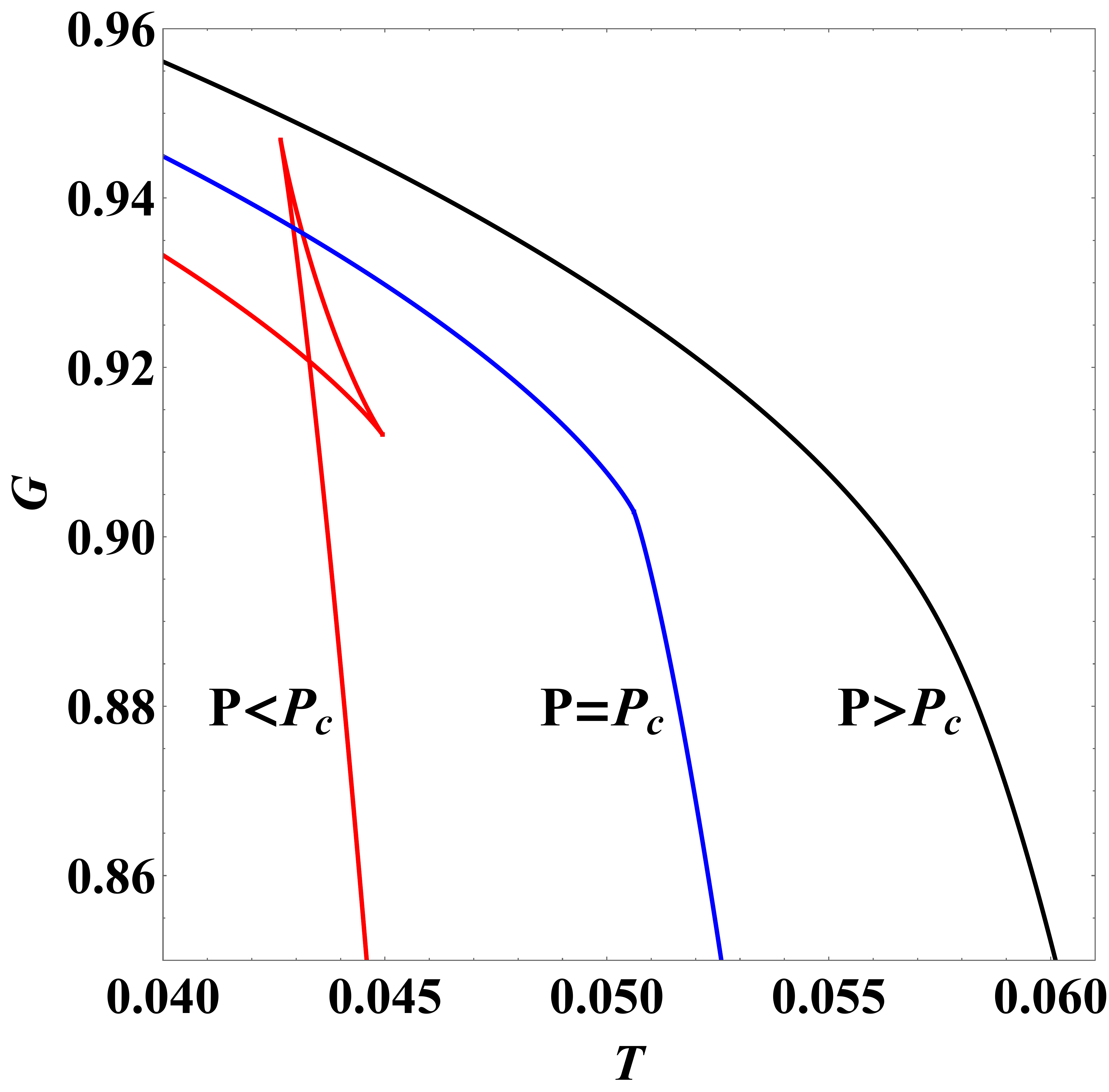}
\label{fig:Fig.5(a)}
\end{minipage}
} 
\subfloat[]
{\begin{minipage}[b]{.33\linewidth}
\centering
\includegraphics[height=5cm]{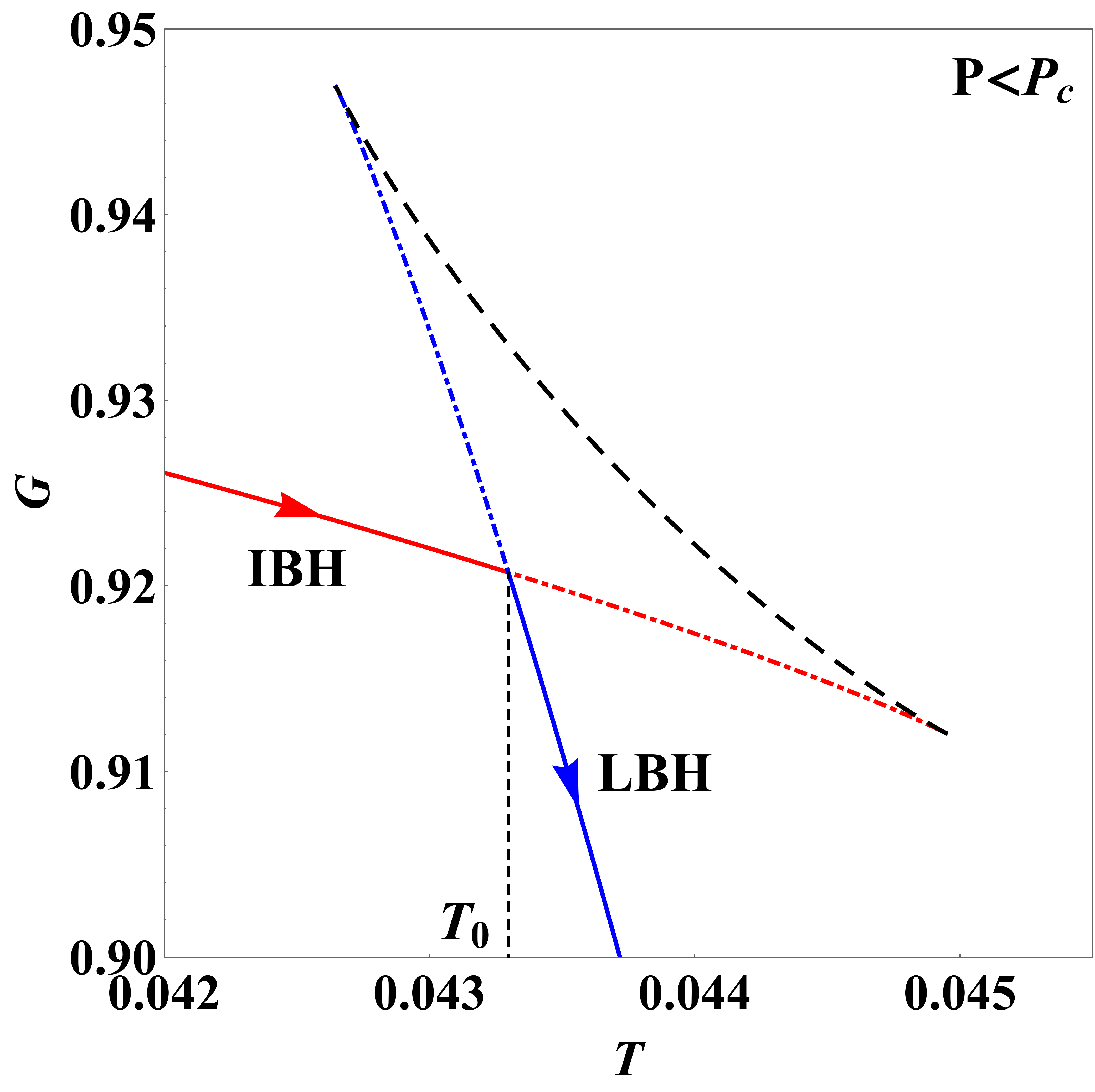}
\label{fig:Fig.5(b)}
\end{minipage}
} 
\subfloat[]
{\begin{minipage}[b]{.33\linewidth}
\centering
\includegraphics[height=5cm]{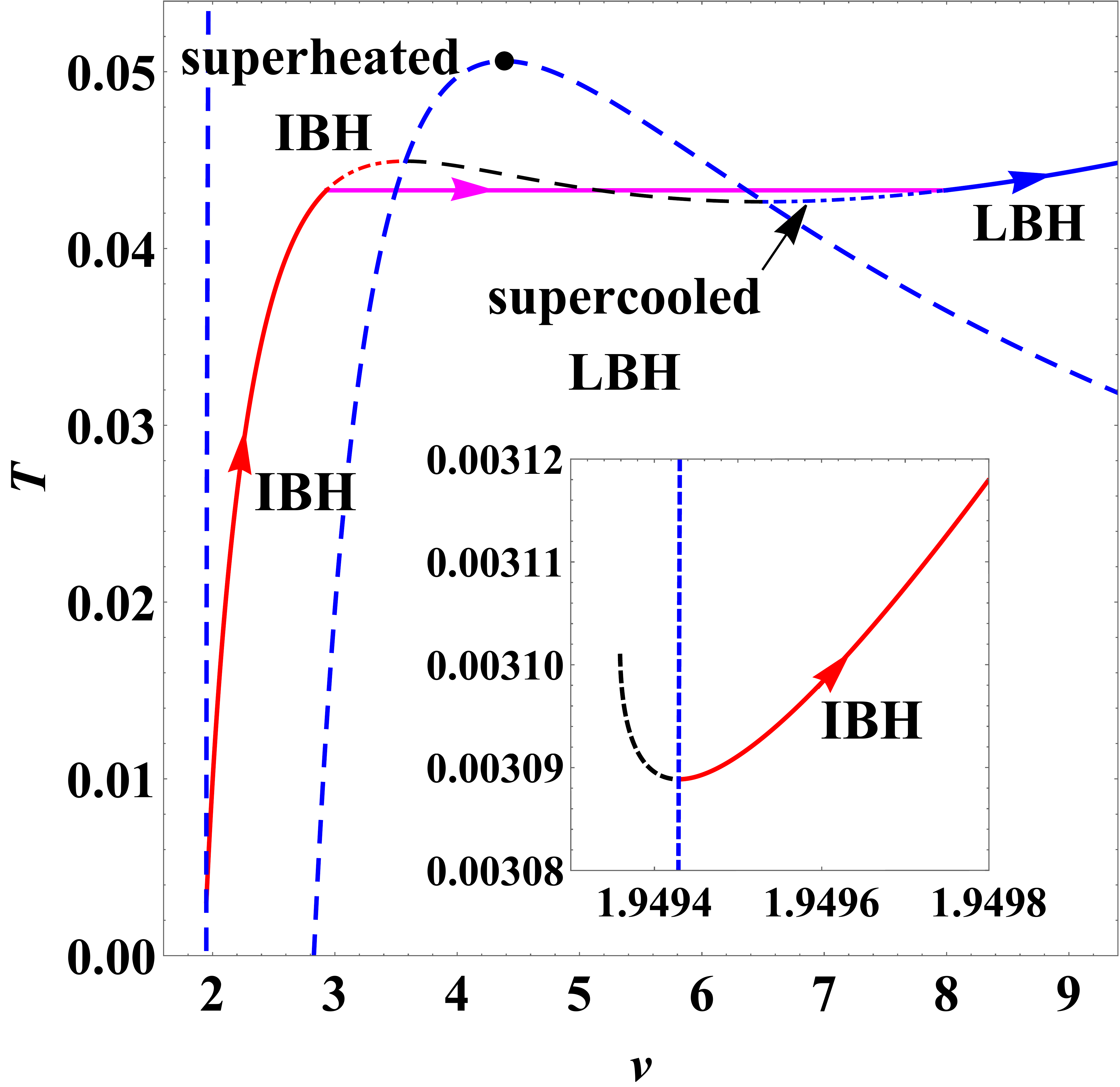}
\label{fig:Fig.5(c)}
\end{minipage}
} 
\caption{\label{fig:Fig.5}(a) The Gibbs free energy for $P = 0.003$, $0.0044$ $(P_{c})$ and $0.006$ from left to right. (b) The VdW-like phase transition between IBH and LBH phases. (c) The isobaric curve for $P = 0.003$. The red and blue solid lines denote the stable IBH and LBH hole branches, respectively. The black hole branches marked with dashed lines are unstable. Black dots represent the critical points. We have set $\alpha^{2}/Q^{2}=0.95$ with $Q = 1$.}
\end{figure}

Here we show the behaviors of the Gibbs free energy for the VdW-like PT case in Fig. \ref{fig:Fig.5}(\subref*{fig:Fig.5(a)}). We take $\alpha^{2}/Q^{2}=0.95$ for example. A typical swallowtail behavior occurs when $P<P_{c}$, which is the signature of VdW-like PT. The swallowtail gets smaller with the increasing of pressure, which turns into a point at the critical pressure $P_{c}$ and eventually disappears beyond $P_{c}$.  

\begin{figure}
\centering
\subfloat[$\alpha^{2}/Q^{2}= 0$]
{\begin{minipage}[b]{.33\linewidth}
\centering
\includegraphics[height=5.07cm]{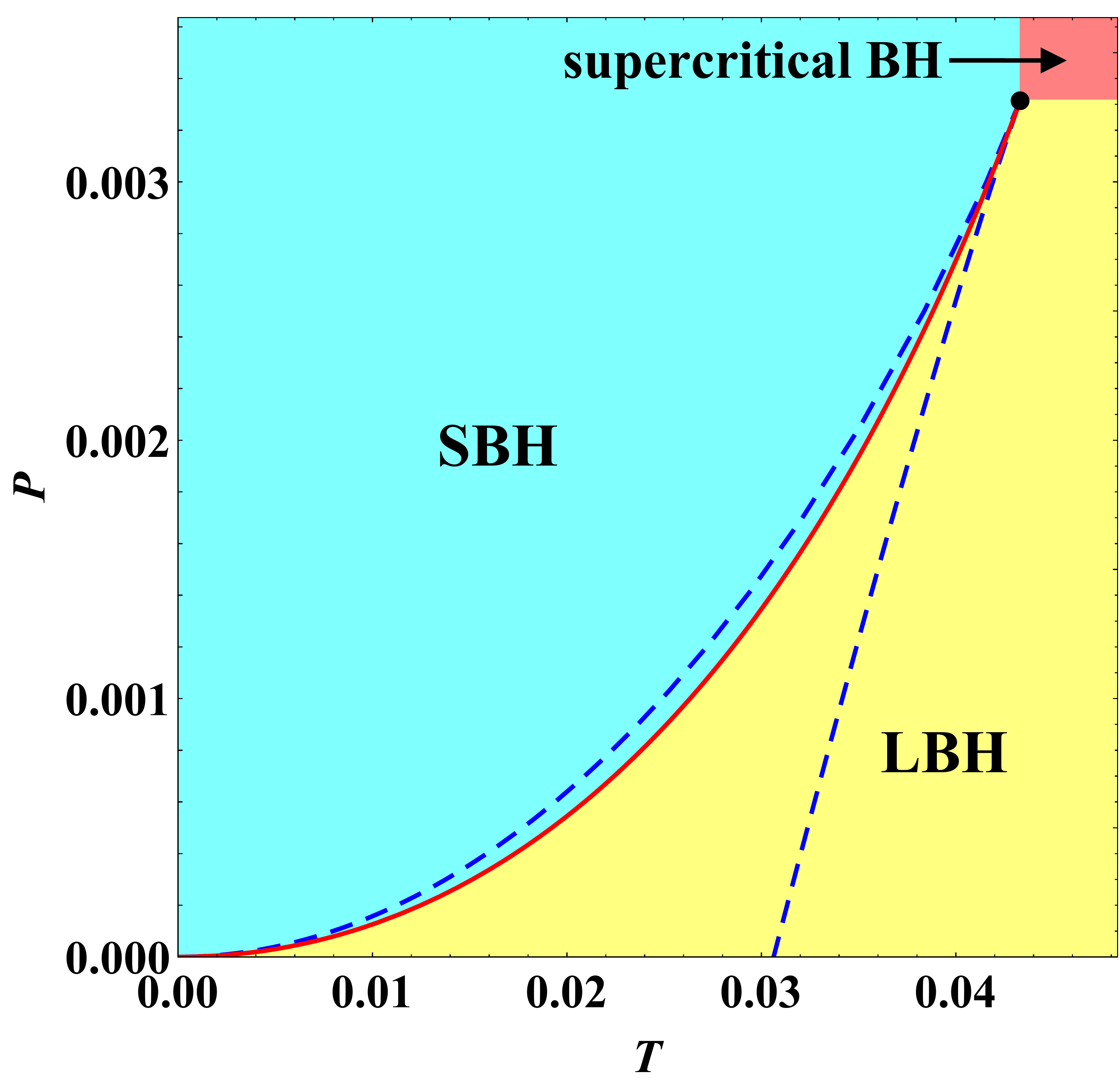}
\label{fig:Fig.6(a)}
\end{minipage}
} 
\subfloat[$\alpha^{2}/Q^{2}=0.95$]
{\begin{minipage}[b]{.33\linewidth}
\centering
\includegraphics[height=5cm]{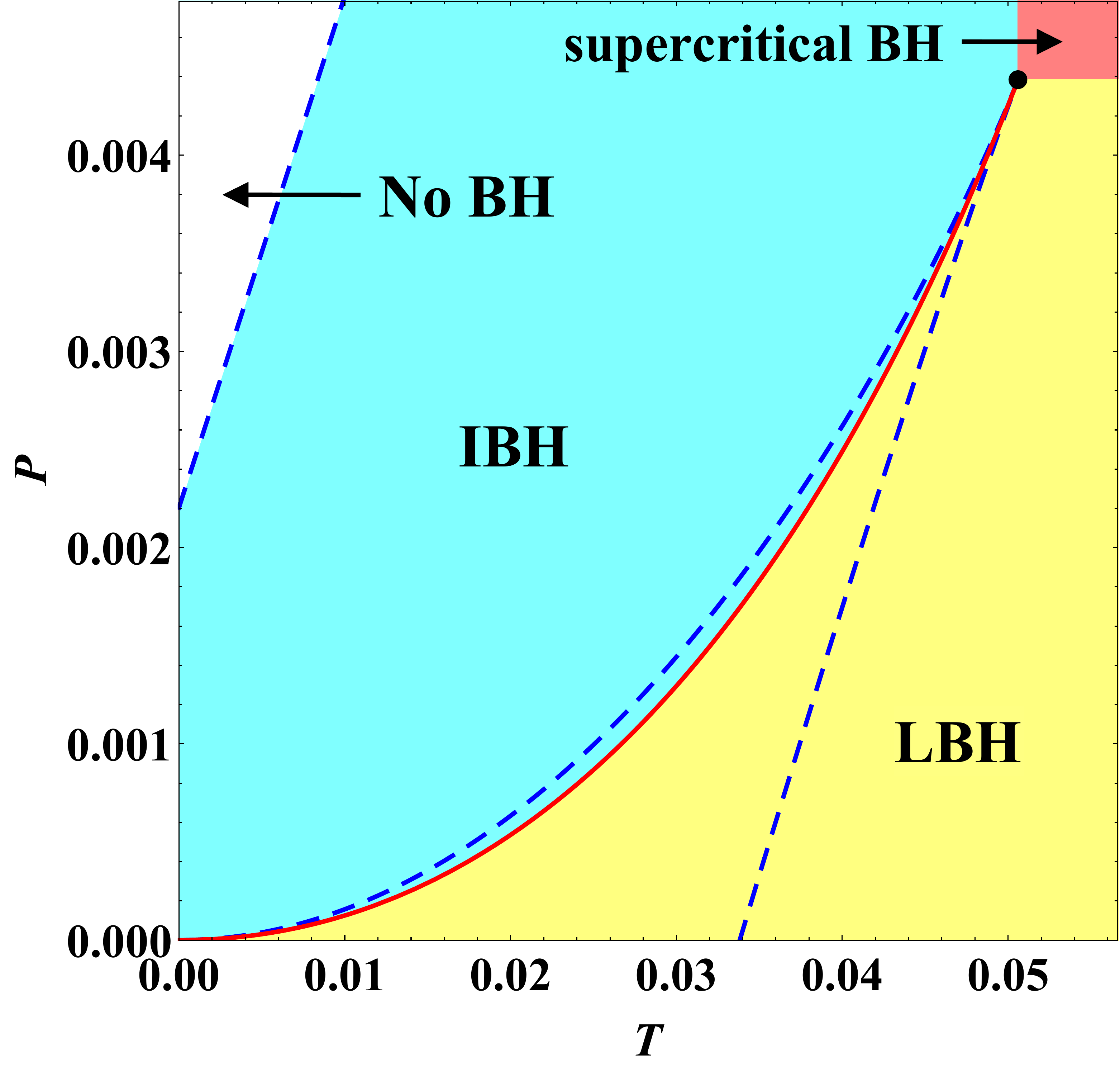}
\label{fig:Fig.6(b)}
\end{minipage}
}
\subfloat[$\alpha^{2}/Q^{2}=1$]
{\begin{minipage}[b]{.33\linewidth}
\centering
\includegraphics[height=5cm]{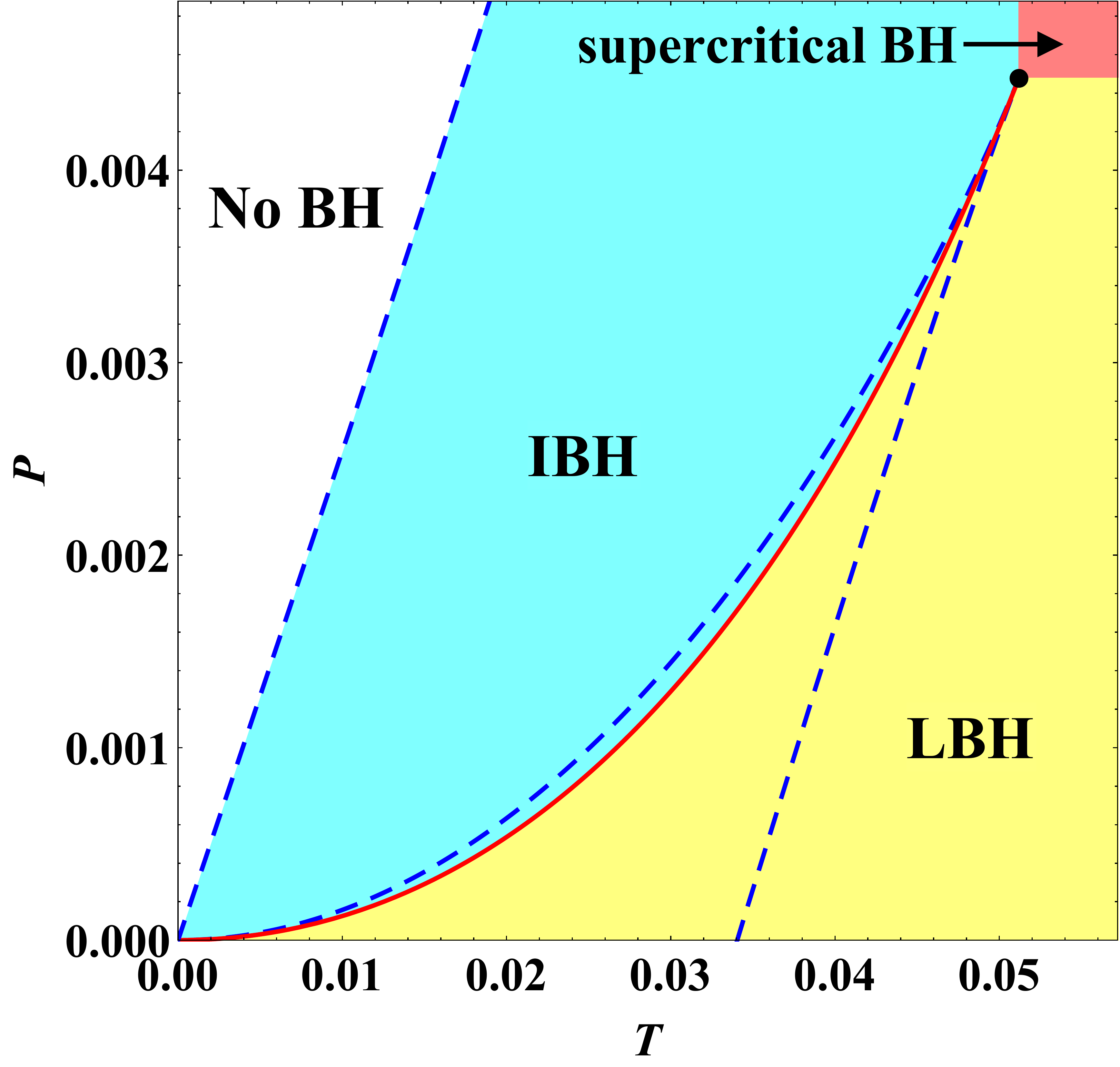}
\label{fig:Fig.6(c)}
\end{minipage}
}
\quad
\subfloat[$\alpha^{2}/Q^{2}= 0$]
{\begin{minipage}[b]{.33\linewidth}
\centering
\includegraphics[height=5cm]{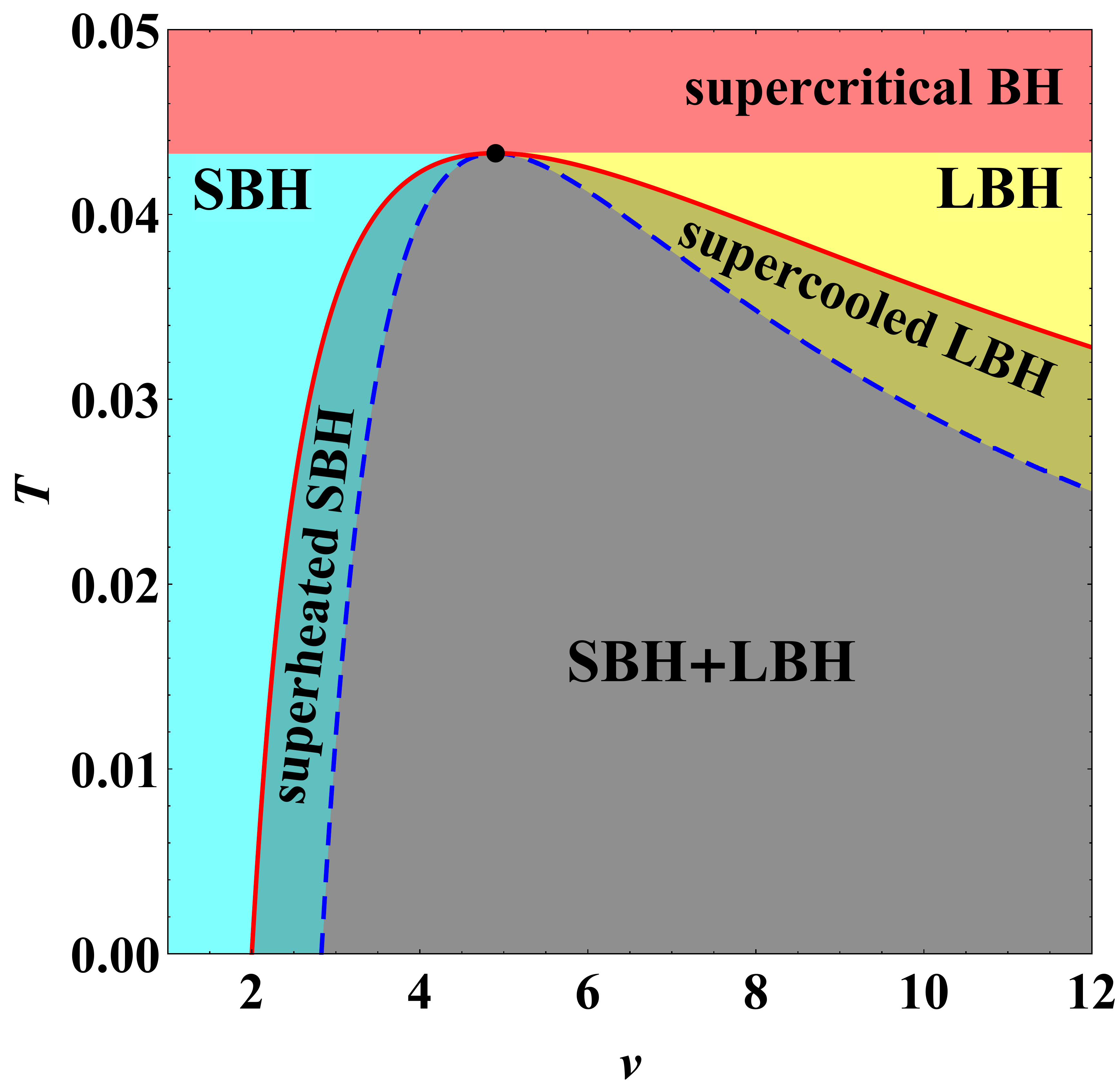}
\label{fig:Fig.6(d)}
\end{minipage}
} 
\subfloat[$\alpha^{2}/Q^{2}=0.95$]
{\begin{minipage}[b]{.33\linewidth}
\centering
\includegraphics[height=5cm]{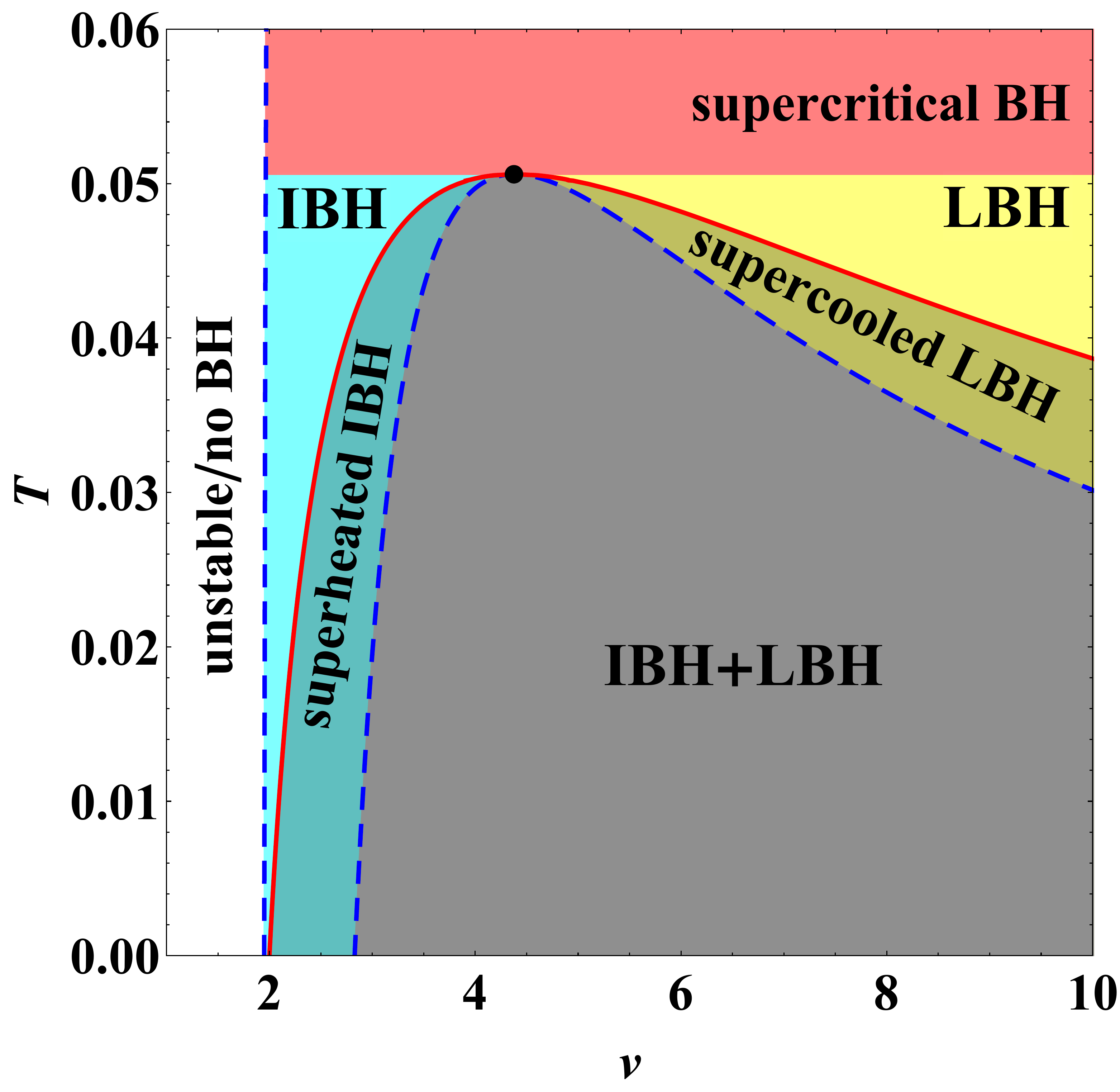}
\label{fig:Fig.6(e)}
\end{minipage}
}
\subfloat[$\alpha^{2}/Q^{2}=1$]
{\begin{minipage}[b]{.33\linewidth}
\centering
\includegraphics[height=5cm]{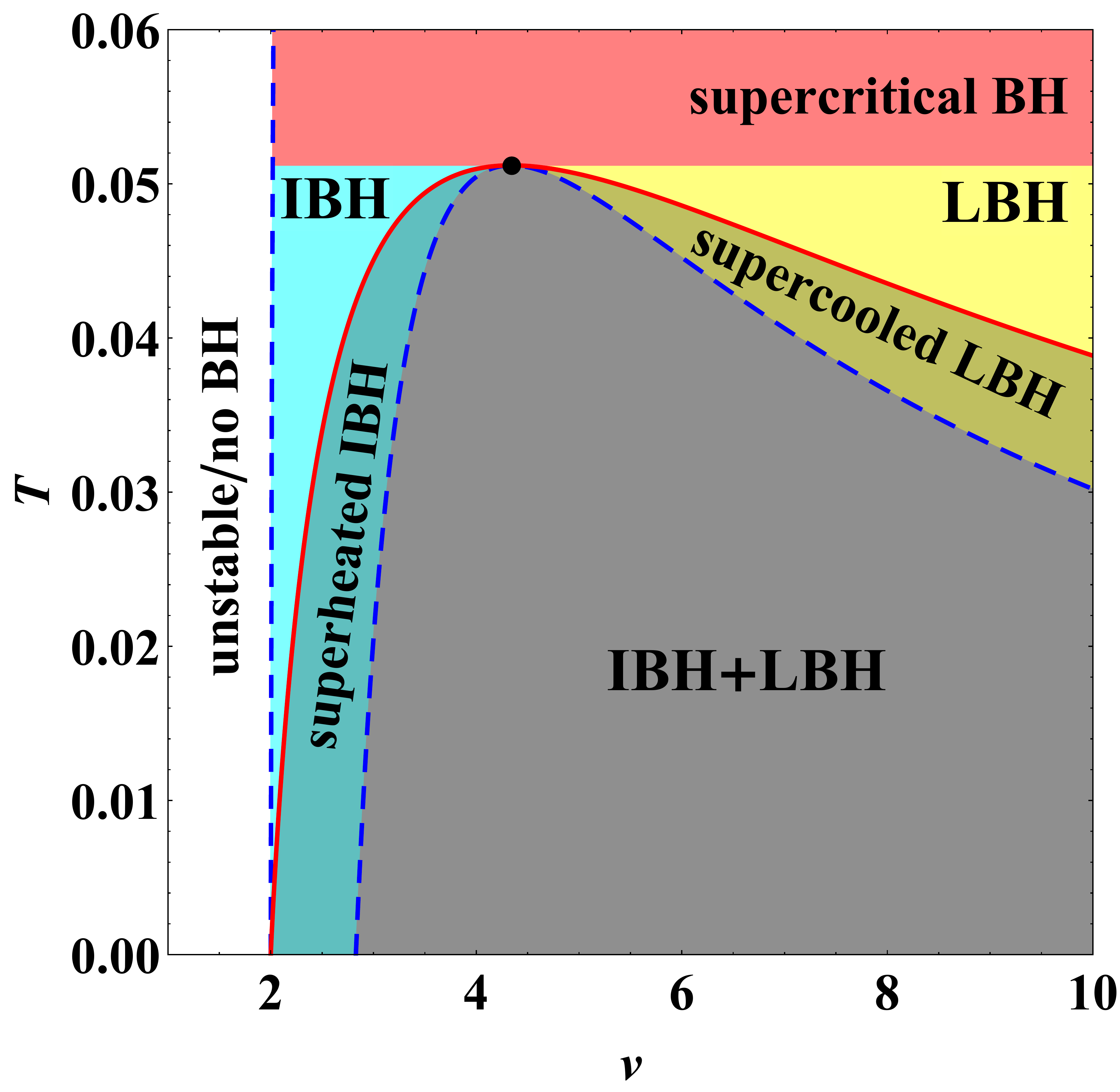}
\label{fig:Fig.6(f)}
\end{minipage}
}
\caption{\label{fig:Fig.6}Phase structures of the GUP-corrected charged AdS black hole for $\alpha^{2}/Q^{2}= 0$, $0.95$ and $1$ in $P-T$ phase diagrams (a-c) and $T-v$ phase diagrams (d-f). The red solid curves represent the coexistence curves. The blue dashed curves are the spinodal curves. Black dots represent the critical points. We have set $Q = 1$.}
\end{figure}

In Figs. \ref{fig:Fig.5}(\subref*{fig:Fig.5(b)}) and \ref{fig:Fig.5}(\subref*{fig:Fig.5(c)}), we describe the behaviors of Gibbs free energy and isobaric curve for $P<P_{c}$ in detail. These branches in solid curves are stable, while the dashed ones are unstable or metastable. Since there is another tiny unstable branch in the small black hole region (see the dashed line in the inset of Fig. \ref{fig:Fig.5}(\subref*{fig:Fig.5(c)})), we would like to call these two stable branches IBH and LBH instead of SBH and LBH. Starting with the lowest temperature, the system stays in the IBH phase (red solid curve) first for its lower Gibbs free energy. At the temperature $T_{0}$, the system transits into the LBH phase, and gains lower Gibbs free energy. Hence the IBH/LBH first-order phase transition occurs at this point, analogous to the liquid-gas phase transition of VdW fluid. Two more metastable branches, including the superheated IBH branch (red dot-dashed curve) and the supercooled LBH branch (blue dot-dashed curve), are also shown in Fig. \ref{fig:Fig.5}(\subref*{fig:Fig.5(c)}). These black holes are locally stable, for they are above the right branch of spinodal curve (blue dashed line). However, they do not have the lower Gibbs free energy. A detailed study about these metastable black holes can be found in Ref. \cite{Wei:2015ana}.

The phase structure of GUP-corrected charged AdS black holes for the $\alpha^{2}/Q^{2}=0.95$ case can be then obtained, as described in Fig. \ref{fig:Fig.6}(\subref*{fig:Fig.6(b)}). As a comparison, we also display the phase structure for $\alpha^{2}/Q^{2}= 0$ and $\alpha^{2}/Q^{2}=1$ in Figs. \ref{fig:Fig.6}(\subref*{fig:Fig.6(a)}) and \ref{fig:Fig.6}(\subref*{fig:Fig.6(c)}) respectively. For the $\alpha^{2}/Q^{2}=0$ case, namely the case for RN-AdS black hole, three black hole phases including SBH, LBH, and supercritical black hole exist in the $P-T$ diagram. The red solid curve in Fig. \ref{fig:Fig.6}(\subref*{fig:Fig.6(a)}) is the coexistence curve of SBH and LBH. Blue dashed curves at top and bottom represent the spinodal curves for SBH and LBH, respectively. These spinodal curves intersect the coexistence curve at critical point (black dot). Beyond the critical point, different phases cannot be clearly distinguished, leading to the supercritical black hole region. For $\alpha^{2}/Q^{2}\neq 0$, the phase structure is similar to the $\alpha^{2}/Q^{2}= 0$ one, but the SBH phase is replaced by the IBH phase for the reason we explained before. Moreover, there is another region of no black hole. This region is above the new appearing spinodal curve, which corresponds to the left branch of spinodal curve for $\alpha^{2}/Q^{2}\neq 0$ (see Fig. \ref{fig:Fig.2}). With the increasing of $\alpha^{2}/Q^{2}$, the area of the no black hole region increases, and the new appearing spinodal curve  intersects the coexistence curve at origin when $\alpha^{2}/Q^{2}=1$, which in fact implies the beginning of RPT case as we will see in the next section.

We show the phase structures in $T-v$ diagrams to clearly describe the coexistence region, see Figs. \ref{fig:Fig.6}(\subref*{fig:Fig.6(d)}--\subref*{fig:Fig.6(f)}). The blue dashed and red solid curves represent the spinodal and coexistence curves in $T-v$ diagrams, respectively. For the case of $\alpha^{2}/Q^{2}=0$, the SBH region is on the left of the coexistence curve, while the LBH region is on the right. Two more metastable phases, the superheated SBH phase and the supercooled LBH phase, are present between the coexistence curve and the spinodal curve. Below the spinodal curve, it is the region of the coexistence SBH and LBH. The supercritical black hole region is above the critical temperature line. For $\alpha^{2}/Q^{2}\neq 0$, the SBH and superheated SBH phases are replaced by the IBH and superheated IBH phases, as shown in Fig. \ref{fig:Fig.5}(\subref*{fig:Fig.5(c)}). Moreover, there is a new region for no black hole ($v<2\alpha$) and the unstable phase in the small black hole region, which is located at the left side of the left branch of spinodal curve.

\subsection{Reentrant Phase Transition}

\begin{figure}
\centering
\subfloat[]
{\begin{minipage}[b]{.49\linewidth}
\centering
\includegraphics[height=7cm]{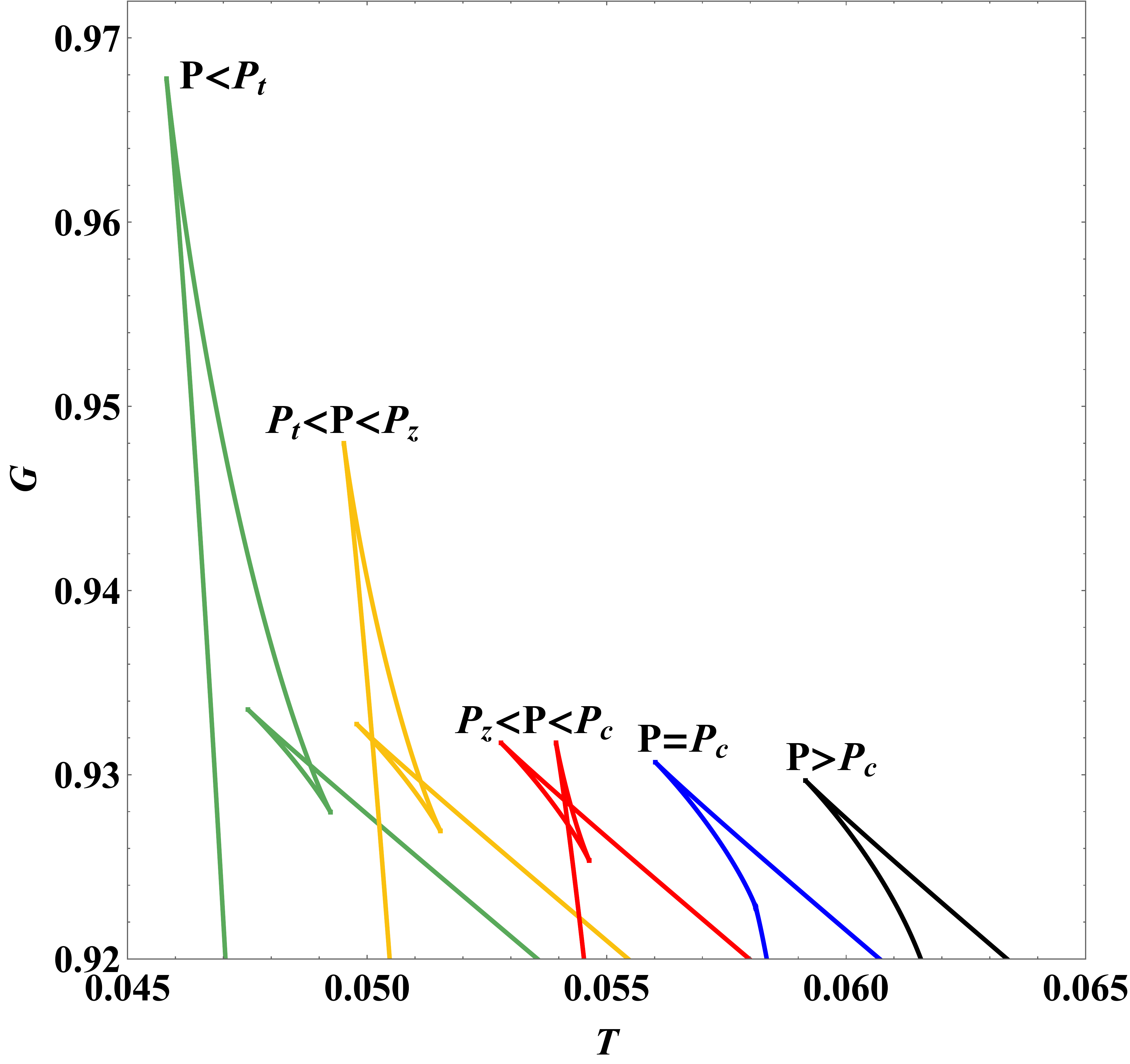}
\label{fig:Fig.7(a)}
\end{minipage}
} 
\subfloat[]
{\begin{minipage}[b]{.49\linewidth}
\centering
\includegraphics[height=7cm]{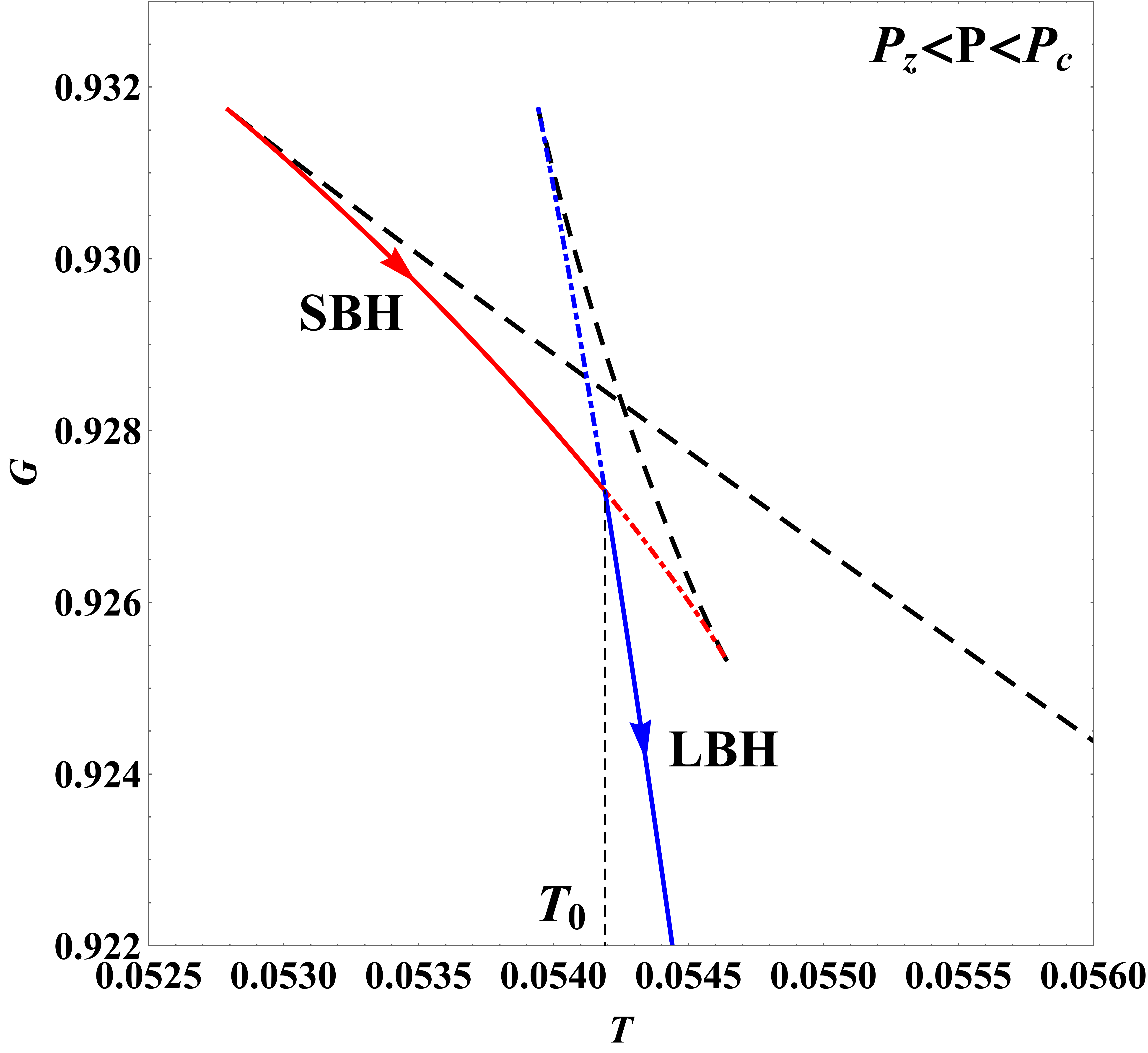}
\label{fig:Fig.7(b)}
\end{minipage}
} 
\quad
\subfloat[]
{\begin{minipage}[b]{.49\linewidth}
\centering
\includegraphics[height=7cm]{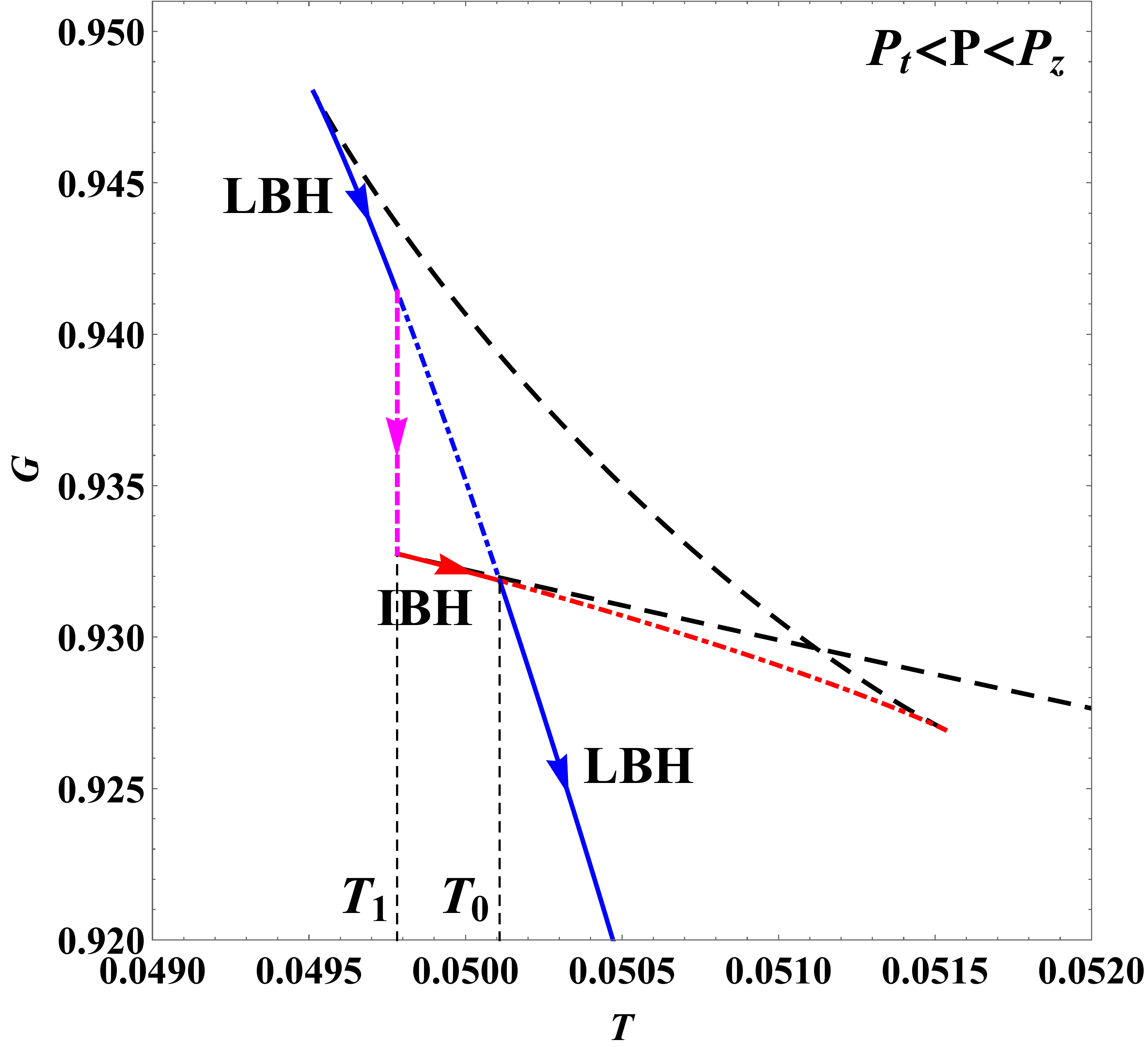}
\label{fig:Fig.7(c)}
\end{minipage}
} 
\subfloat[]
{\begin{minipage}[b]{.49\linewidth}
\centering
\includegraphics[height=7cm]{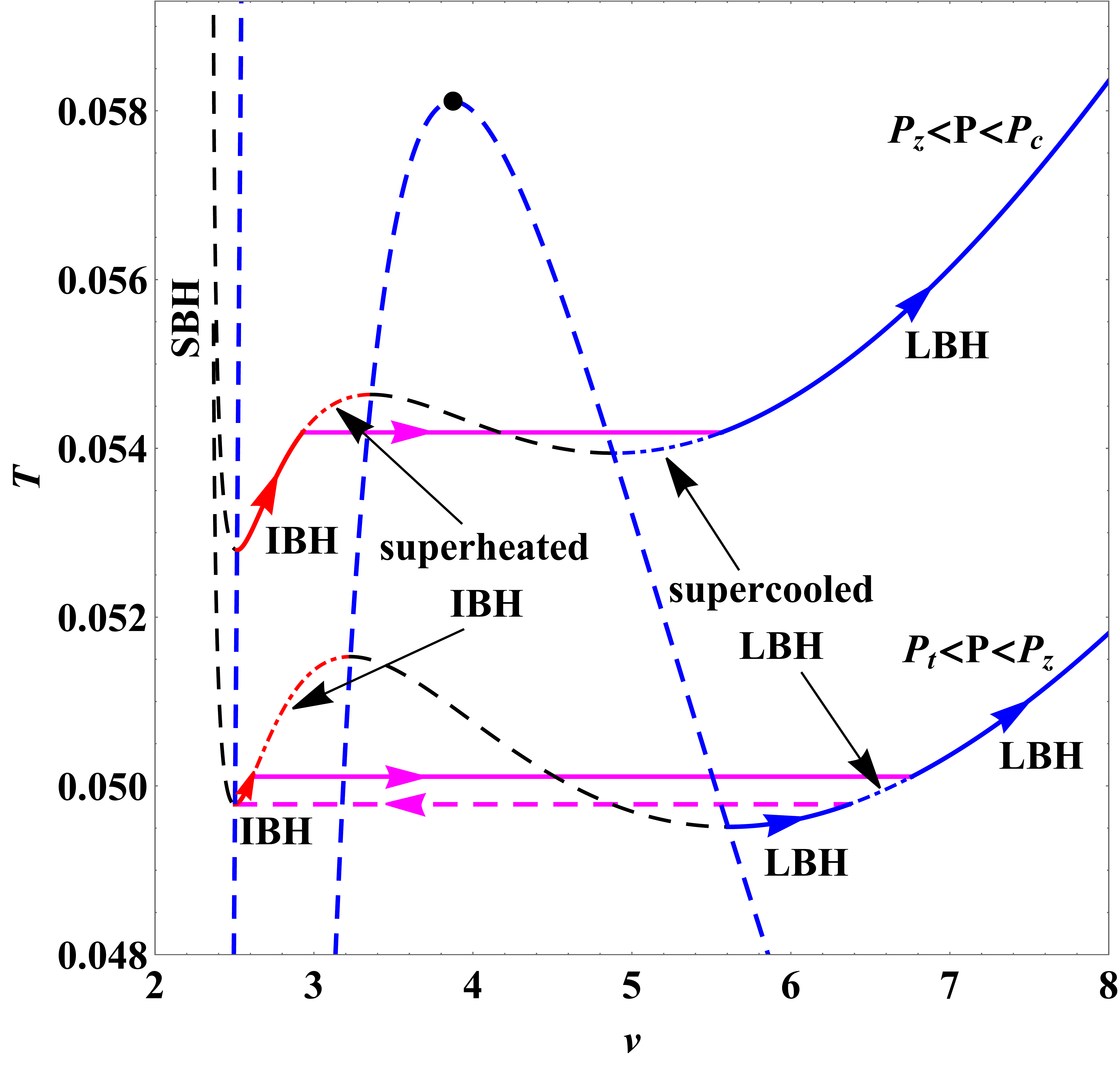}
\label{fig:Fig.7(d)}
\end{minipage}
} 
\caption{\label{fig:Fig.7}(a) The Gibbs free energy for $P = 0.0034, 0.004, 0.0048, 0.0057$ $(P_{c})$ and $0.0065$ from left to right. $P_{t}=0.0038$ and $P_{z}=0.0041$. (b) The VdW-like phase transition between IBH and LBH phase for $P=0.0048$. (c) RPT (LBH/IBH/LBH phase transition) for $P=0.004$. (d) Isobaric curves for $P=0.004$ ($P_{t}<P<P_{z}$) and $0.0048$ ($P_{z}<P<P_{c}$). The red and blue solid lines denote the stable IBH and LBH branches, respectively. The black hole branches marked with dashed lines are unstable. Black dots represent the critical points. We have set $\alpha^{2}/Q^{2}=1.4$ with $Q = 1$.}
\end{figure}

The case of the RPT becomes complicated. Three cases of RPT are displayed by the black hole, including $1<\alpha^{2}/Q^{2}<1.5$, $\alpha^{2}/Q^{2}=1.5$ and $1.5<\alpha^{2}/Q^{2}<1.535$. Taking $\alpha^{2}/Q^{2}=1.4$ for example, we depict the Gibbs free energy with different pressures in Fig. \ref{fig:Fig.7}(\subref*{fig:Fig.7(a)}). For $P_{z}<P<P_{c}$ the behavior of the system is similar to the IBH/LBH first-order phase transition of the VdW-like case, see Fig. \ref{fig:Fig.7}(\subref*{fig:Fig.7(b)}) and the corresponding isobaric curve in Fig. \ref{fig:Fig.7}(\subref*{fig:Fig.7(d)}). Increasing the pressure, we find an additional second-order phase transition at $T=T_{c}$ when $P=P_{c}$, and there will be no more phase transition when $P>P_{c}$. In the region $P_{t} < P < P_{z}$, a zeroth-order phase transition occurs besides the IBH/LBH first-order phase transition (see Figs. \ref{fig:Fig.7}(\subref*{fig:Fig.7(c)}) and \ref{fig:Fig.7}(\subref*{fig:Fig.7(d)})). Starting with the lowest temperature of the Gibbs free energy curve and the corresponding isobaric curve, the black hole evolves along the LBH branch first. Then at some point $T_{1}$ the black hole transits to the IBH branch by a zeroth-order phase transition. Upon reaching temperature $T_{0}$, the black hole undergoes a first-order phase transition and eventually reenters the LBH branch. This behavior of the Gibbs free energy indicates a typical RPT. Two metastable branches namely the superheated IBH and supercooled LBH are also clearly shown in Fig. \ref{fig:Fig.7}(\subref*{fig:Fig.7(d)}). It is noteworthy that the supercooled LBH is between the two branches of LBH when $P_{t} < P < P_{z}$. For $P < P_{t}$, the situation is the same as the $P > P_{c}$ case, where the LBH branch always has the lower Gibbs free energy.

\begin{figure}
\centering
\subfloat[$\alpha^{2}/Q^{2}=1.4$]
{\begin{minipage}[b]{.33\linewidth}
\centering
\includegraphics[height=4.9cm]{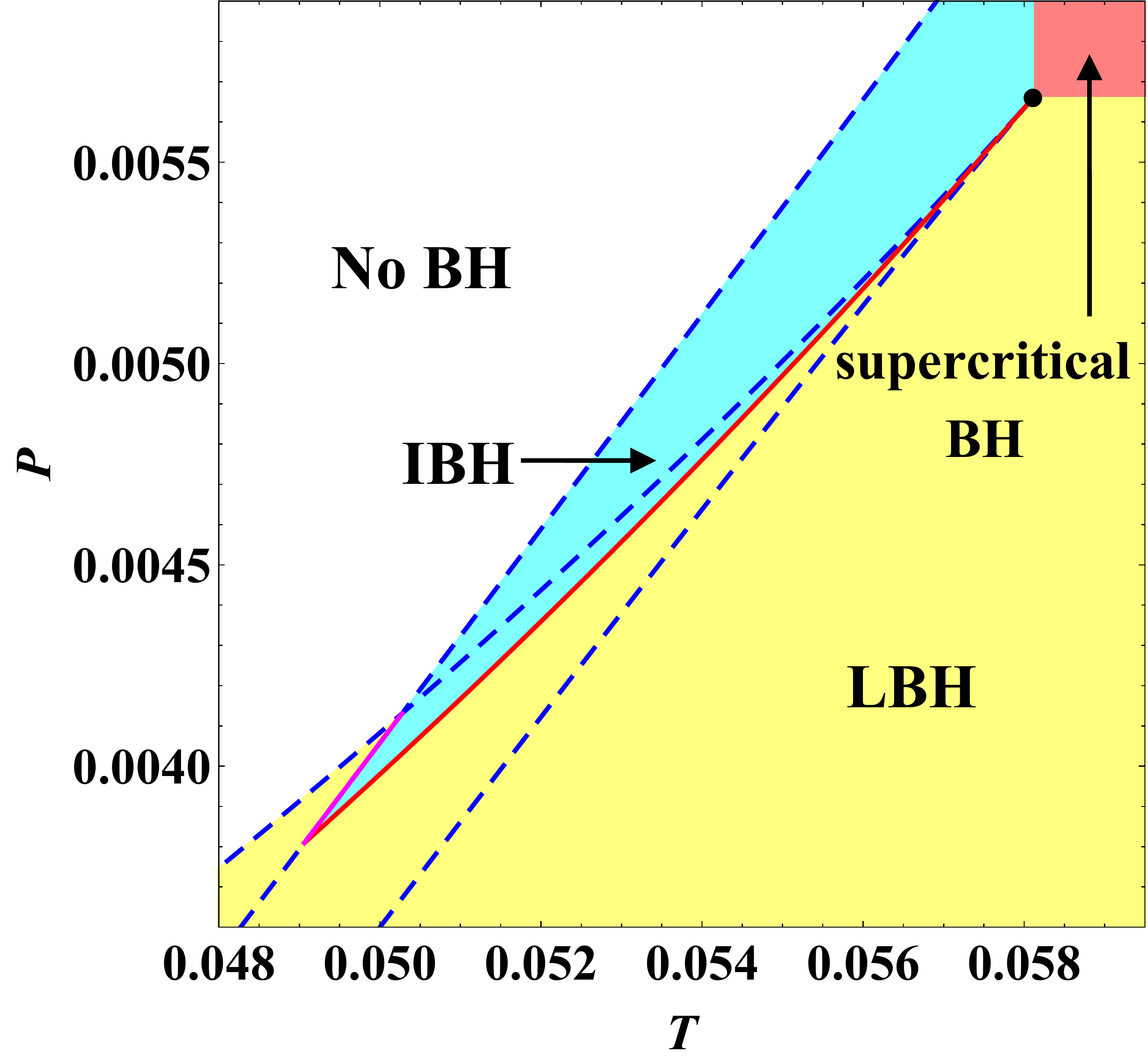}
\label{fig:Fig.8(a)}
\end{minipage}
} 
\subfloat[$\alpha^{2}/Q^{2}=1.5$]
{\begin{minipage}[b]{.33\linewidth}
\centering
\includegraphics[height=4.9cm]{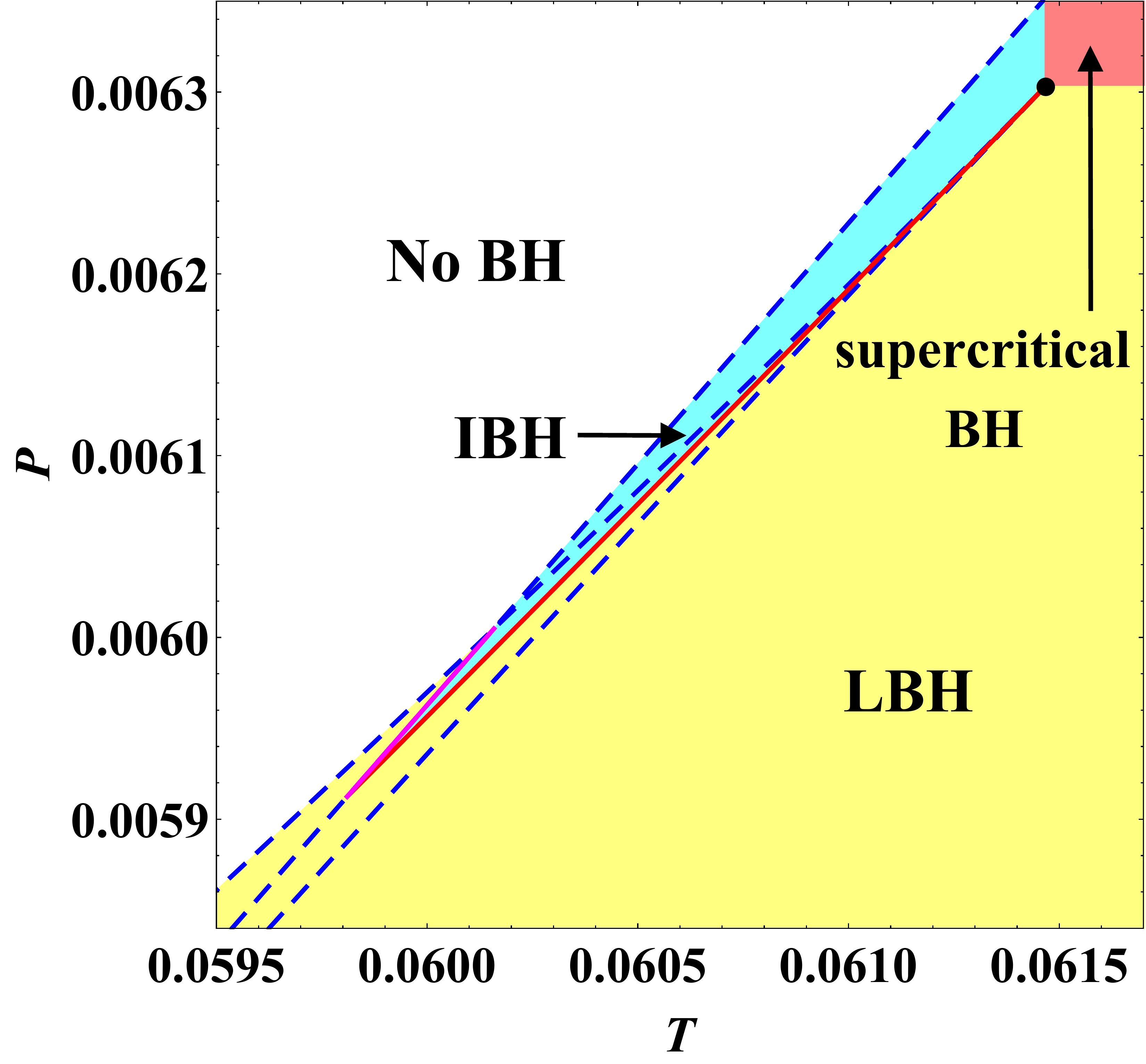}
\label{fig:Fig.8(b)}
\end{minipage}
}
\subfloat[$\alpha^{2}/Q^{2}=1.51$]
{\begin{minipage}[b]{.33\linewidth}
\centering
\includegraphics[height=5cm]{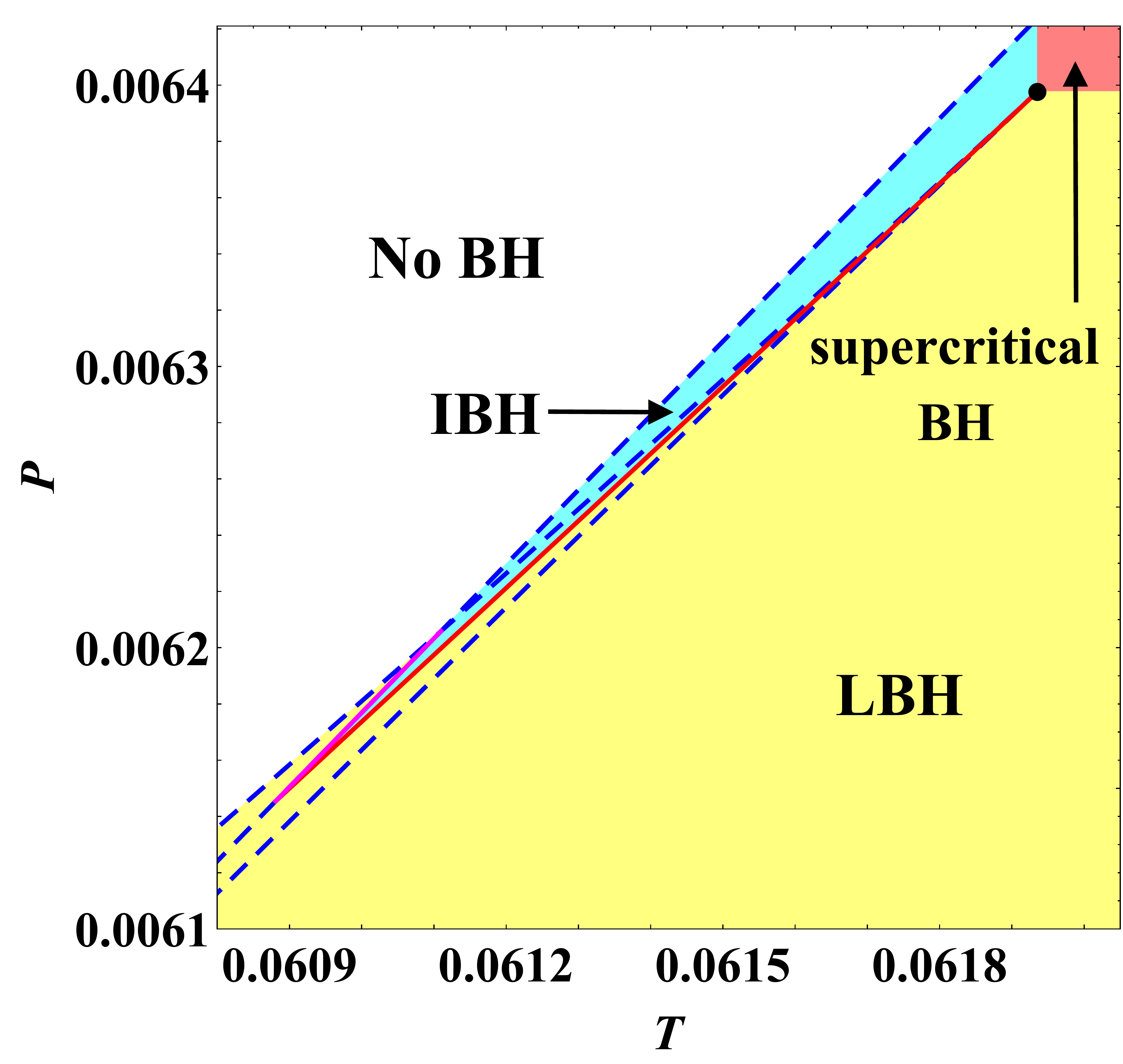}
\label{fig:Fig.8(c)}
\end{minipage}
}
\quad
\subfloat[$\alpha^{2}/Q^{2}=1.4$]
{\begin{minipage}[b]{.33\linewidth}
\centering
\includegraphics[height=4.9cm]{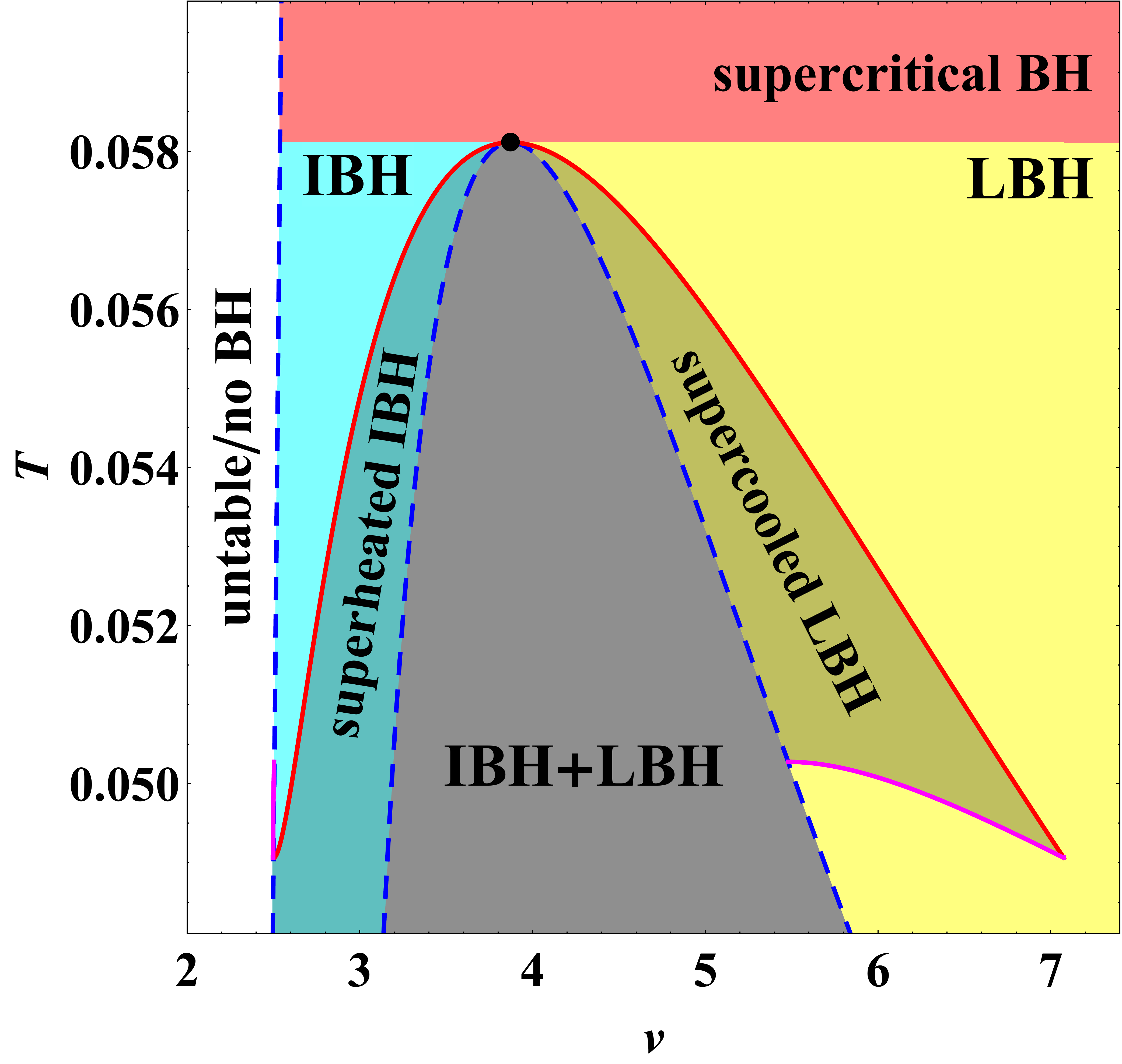}
\label{fig:Fig.8(d)}
\end{minipage}
} 
\subfloat[$\alpha^{2}/Q^{2}=1.5$]
{\begin{minipage}[b]{.33\linewidth}
\centering
\includegraphics[height=4.9cm]{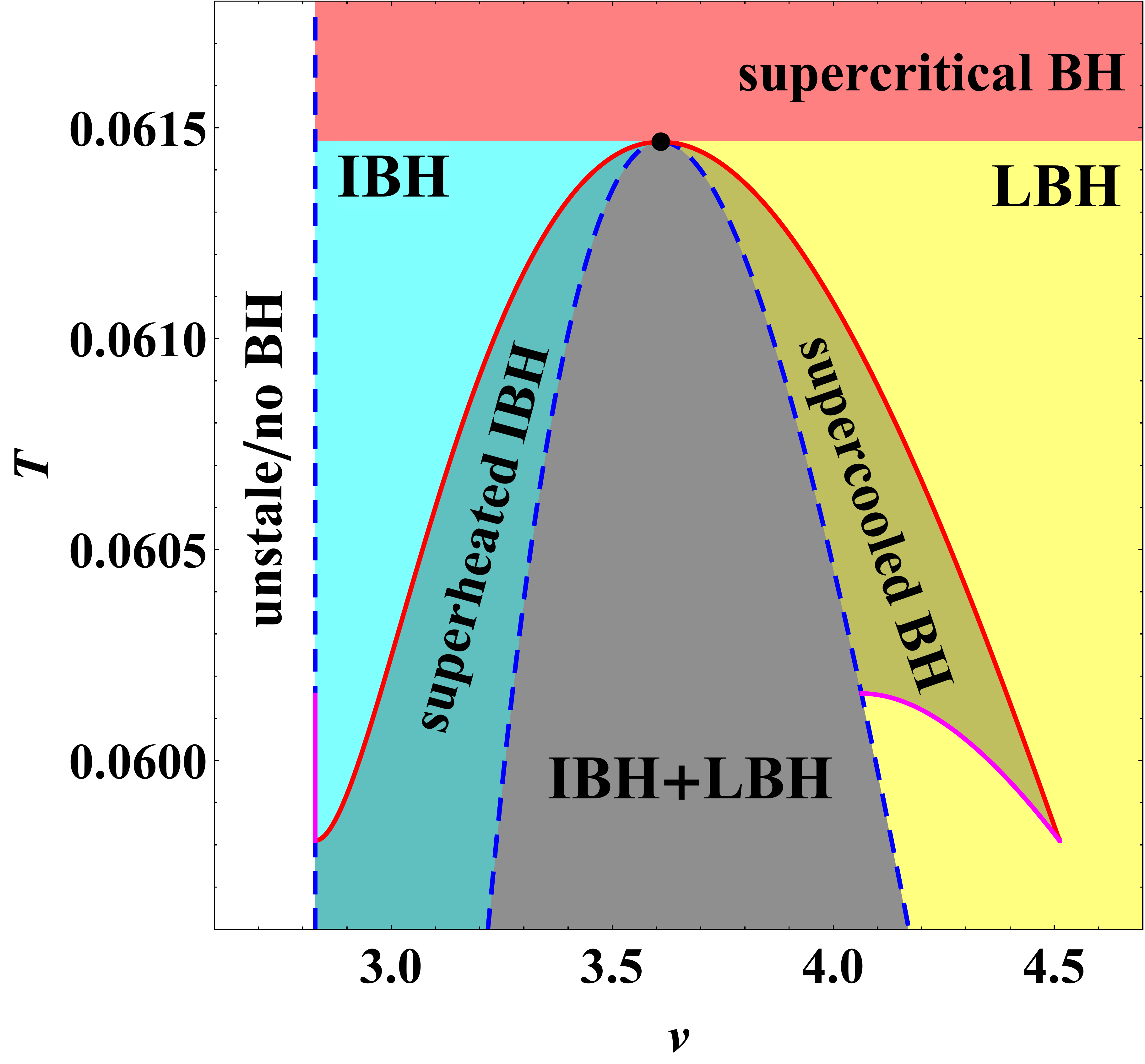}
\label{fig:Fig.8(e)}
\end{minipage}
}
\subfloat[$\alpha^{2}/Q^{2}=1.51$]
{\begin{minipage}[b]{.33\linewidth}
\centering
\includegraphics[height=4.9cm]{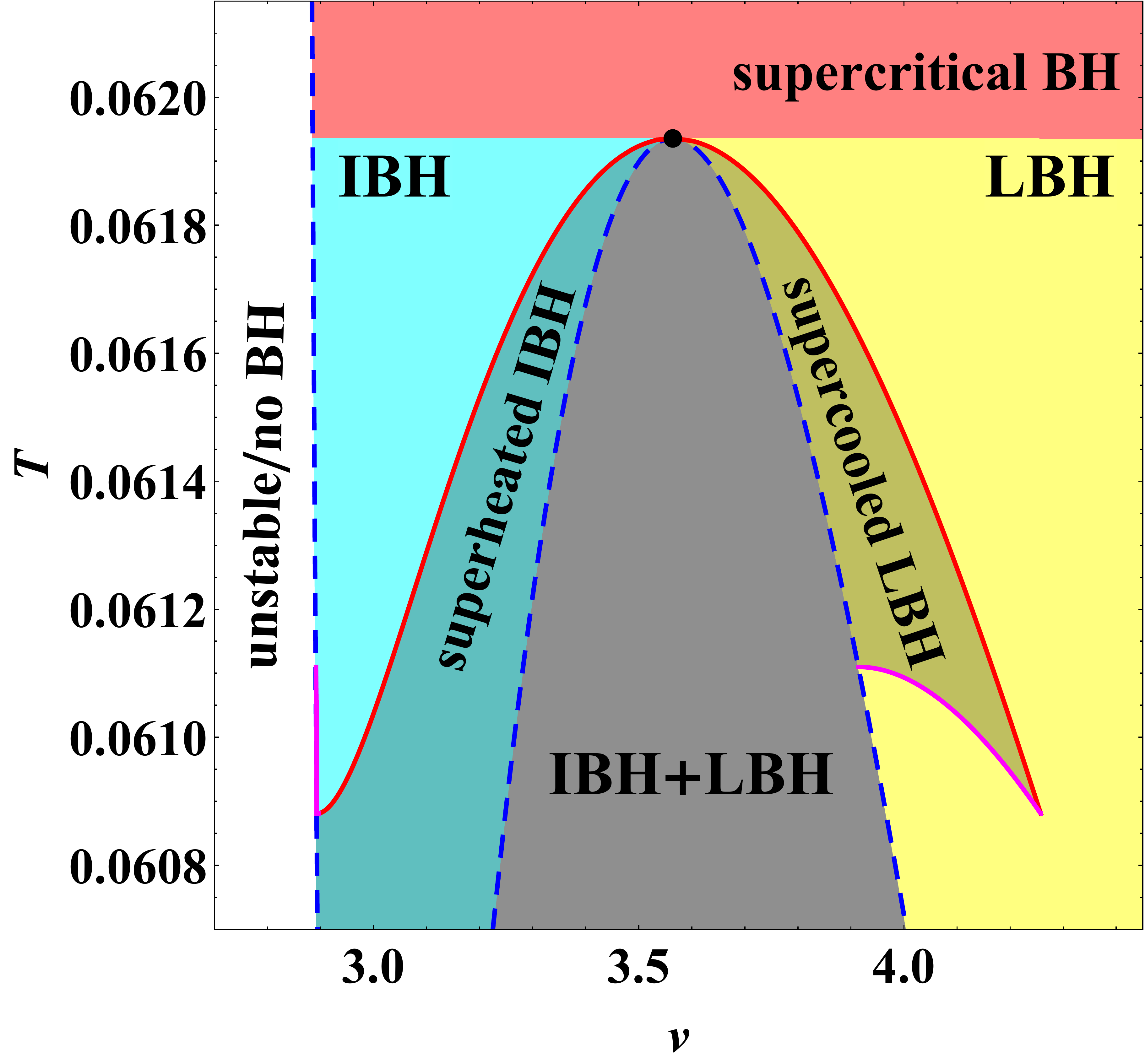}
\label{fig:Fig.8(f)}
\end{minipage}
}
\caption{\label{fig:Fig.8}Phase structures of the GUP-corrected charged AdS black hole for $\alpha^{2}/Q^{2}=1.4$, $1.5$ and $1.51$ in $P-T$ phase diagrams (a-c) and $T-v$ phase diagrams (d-f). The red solid curves represent the coexistence curves. The magenta solid curves are for the zeroth-order phase transition curves. The spinodal curves are marked with blue dashed curves. Black dots represent the critical points. We have set $Q = 1$.}
\end{figure}

We demonstrate the phase structures for the $\alpha^{2}/Q^{2}=1.4$ case in the $P-T$ and $T-v$ diagrams, see Figs. \ref{fig:Fig.8}(\subref*{fig:Fig.8(a)}) and \ref{fig:Fig.8}(\subref*{fig:Fig.8(c)}) respectively. Red and magenta solid lines depict the first-order (coexistence curve) and zeroth-order phase transition curves of the IBH and LBH, respectively. The spinodal curves are represented here too as blue dashed curves. In the $P-T$ diagram, the upper left corner bounded by two of these spinodal curves is the region without any black hole. It is noteworthy that the zeroth-order phase transition curve overlaps with the spinodal curve for $P < P_{z}$ and $T < T_{z}$, which can be confirmed by the zeroth-order phase transition points corresponding to the intersection points of the isobaric curves of IBHs and the unstable black holes in the small black hole region, i.e., extremal points of the isobaric curves (see Figs. \ref{fig:Fig.7}(\subref*{fig:Fig.7(c)}) and \ref{fig:Fig.7}(\subref*{fig:Fig.7(d)})). In the $T-v$ diagram, two additional metastable phases are clearly displayed, including the superheated IBH phase and the supercooled LBH phase. Comparing to the VdW-like PT case, there is another region for the LBH under the supercooled LBH region, as we discussed before.

For $\alpha^{2}/Q^{2}=1.5$, the phase structures in the $P-T$ and $T-v$ diagrams are shown in Figs. \ref{fig:Fig.8}(\subref*{fig:Fig.8(b)}) and \ref{fig:Fig.8}(\subref*{fig:Fig.8(e)}), respectively. These phase diagrams are analogous to the $\alpha^{2}/Q^{2}=1.4$ case, but the position of stable phases is different. In particular, in the $T-v$ diagram, these stable phases shift to the right with the increasing of $\alpha$, as explained before. Moreover, there are two critical points $c$ and $c_{1}$ in the system when $\alpha^{2}/Q^{2}=1.5$ (see Fig. \ref{fig:Fig.3}(\subref*{fig:Fig.3(b)})). However, the left one $c_{1}$ has no physical significance for there is no phase transition, and thus we do not show it in the phase diagrams.

For $1.5<\alpha^{2}/Q^{2}<1.535$, this is another case of RPT shown by the black hole. Taking $\alpha^{2}/Q^{2}=1.51$ for example, we present the phase structures in the $P-T$ and $T-v$ diagrams in Figs. \ref{fig:Fig.8}(\subref*{fig:Fig.8(c)}) and \ref{fig:Fig.8}(\subref*{fig:Fig.8(f)}), which are similar to the RPT cases discussed above. Here we only show the phases on the right side of line $v=4\sqrt{3} \alpha /3$, because the locally stable phase under the left branch of spinodal curve (see Fig. \ref{fig:Fig.3}(\subref*{fig:Fig.3(c)})), is globally unstable due to a higher Gibbs free energy.

\section{Ruppeiner geometry for GUP-corrected charged AdS black holes}\label{section4}

Now we explore the microstructure of the GUP-corrected charged AdS black hole by use of Ruppeiner geometry. In Ruppeiner geometry, the fluctuation of entropy $\Delta S$ is associated with the line element in the Riemannian geometry \cite{Ruppeiner:1995zz}
\begin{equation}
\label{eq:24}
\Delta l^{2}=-\frac{\partial^{2} S}{\partial x^{\mu} \partial x^{\nu}} \Delta x^{\mu} \Delta x^{\nu},
\end{equation}
where $x^{\mu}$ are the independent thermodynamic quantities. Within this framework, empirical results reveal that the scalar curvature $R$ obtained from Eq. (\ref{eq:24}) relates to the microscopic interaction in a thermodynamic system \cite{ruppeiner1979thermodynamics,ruppeiner1981application,janyszek1989riemannian,janyszek1990riemannian,oshima1999riemann,mirza2009nonperturbative,may2013thermodynamic}. Specially, the positive (negative) $R$ implies the repulsive (attractive) interaction, whereas $R=0$ corresponds to no interaction.

Taking the first law of standard thermodynamics to be the most basic differential relation of thermodynamic quantities, the line element in the $(T,V)$ phase space can be shown as \cite{Wei:2019uqg,Wei:2019yvs}
\begin{equation}
\label{eq:25}
\Delta l^{2}=\frac{C_{V}}{T^{2}} \Delta T^{2}-\frac{\left(\partial_{V} P\right)_{T}}{T} \Delta V^{2},
\end{equation}
where $C_{V}$ is the heat capacity at constant volume
\begin{equation}
\label{eq:26}
C_{V}=T\left(\frac{\partial S}{\partial T}\right)_{V}.
\end{equation}

For the charged AdS black hole, the heat capacity $C_{V}$ vanishes because of the interdependence of volume and entropy, resulting in a pathological scalar curvature derived from the Eq. (\ref{eq:25}). To address this problem, we can eliminate $C_{V}$ by normalizing the scalar curvature as \cite{Wei:2019uqg,Wei:2019yvs}
\begin{equation}
\label{eq:27}
R_{N}=C_{V} R,
\end{equation}
and then we explore the microstructure of the black hole with $R_{N}$. For a state equation of general charged AdS black hole expressed as Eq. (\ref{eq:16}), the normalised scalar curvature can be obtained as
\begin{equation}
\label{eq:28}
R_{N}=\frac{\left(\partial_{V}b\right)_{Q}\left[2T\partial_{V}a+\left(\partial_{v}b\right)_{Q}\right]}{2\left[T\partial_{V}a+\left(\partial_{V}b\right)_{Q}\right]^2},
\end{equation}
where $a=a(V)$ and $b=b(V,Q)$ are the coefficients labeled in Eq. (\ref{eq:16}). Obviously, the denominator of $R_{N}$ is non-negative, vanishing when the system approaches the spinodal curve
\begin{equation}
\label{eq:29}
T_{sp}=-\frac{\left(\partial_{V}b\right)_{Q}}{\partial_{V}a}.
\end{equation}
When the numerator of $R_{N}$ equals zero, we get two sign-changing curves
\begin{equation}
\label{eq:30}
T=\frac{T_{sp}}{2}=-\frac{\left(\partial_{V}b\right)_{Q}}{2\partial_{V}a},\quad \left(\partial_{V}b\right)_{Q}=0.
\end{equation}
The first sign-changing curve can be eliminated considering the fact that the state equation is invalid under the coexistence curves \cite{Wei:2019uqg,Wei:2019yvs}. As a result, only the sign-changing curve at $\left(\partial_{V}b\right)_{Q}=0$ has physical significance.

It is worth noting that, by taking the first law of black hole thermodynamics to be the most basic differential relation of thermodynamic quantities, a different line element on the $(T,V)$ has also been given by \cite{Ghosh:2019pwy,Xu:2020gud}
\begin{equation}
\label{eq:31}
\Delta l^{2}=\frac{C_{V}}{T^{2}} \Delta T^{2}+\frac{2}{T}\left(\frac{\partial P}{\partial T}\right)_{V} \Delta T \Delta V+\frac{1}{T}\left(\frac{\partial P}{\partial V}\right)_{T} \Delta V^{2},
\end{equation}
as it is valid even for $C_{V}=0$. Considering the state equation Eq. (\ref{eq:16}), the scalar curvature turns out to be
\begin{equation}
\label{eq:32}
R=-\frac{\left(\partial_{V}b\right)_{Q}}{Ta^2}.
\end{equation}
The denominator of $R$ is non-negative, thus the sign-changing curve is given as
\begin{equation}
\label{eq:33}
\left(\partial_{V}b\right)_{Q}=0.
\end{equation}
It is interesting that the two different line elements Eq. (\ref{eq:25}) and Eq. (\ref{eq:31}) give the same sign-changing curve. In fact, this result can be qualitatively understood using the following equation \cite{Ghosh:2019pwy}
\begin{equation}
\label{eq:34}
\left(\frac{\partial P}{\partial V}\right)_{T=0}=0.
\end{equation}
Since there is no kinetic pressure for $T=0$, the entire pressure results from interactions; therefore, we can treat Eq. (\ref{eq:34}) as the point where the repulsion-dominated black hole turns into the attraction-dominated black hole. However, such a point is not a phase transition point as the thermal effects are not included.

In the following sections, we would like to apply the scalar curvature Eq. (\ref{eq:32}) to probe the microstructure of GUP-corrected charged AdS black hole. The scalar curvature associated with the GUP-corrected state equation Eq. (\ref{eq:14}) yields as
\begin{equation}
\label{eq:35}
R_{GUP}=\frac{4\left[2 Q^{2}-\left(\frac{3}{4 \pi}\right)^{\frac{2}{3}} V^{\frac{2}{3}}\right]}{3 \pi T V\left[\left(\frac{3}{4 \pi}\right)^{\frac{1}{3}} V^{\frac{1}{3}}+\sqrt{\left(\frac{3}{4 \pi}\right)^{\frac{2}{3}} V^{\frac{2}{3}}-\alpha^{2}}\right]^{2}}.
\end{equation}
When the correction for GUP disappears ($\alpha$ = 0), we have
\begin{equation}
\label{eq:36}
R=\frac{\left(4 \sqrt[3]{6} \pi^{2 / 3} Q^{2}\right) / V^{2 / 3}-3}{9 \pi T V},
\end{equation}
which is consistent with the result given in Ref. \cite{Ghosh:2019pwy}. At the first sight, $R_{GUP}$ changes the sign when
\begin{equation}
\label{eq:37}
V=\frac{8\sqrt{2} \pi Q}{3}, \quad \text { or equally}, \quad v=2\sqrt{2}Q,
\end{equation}
which is the same as the result of no GUP correction. As we discussed before, for a general charged AdS black hole, the GUP correction will not influence the $b(V, Q)$ term in the state equation Eq. (\ref{eq:16}). Meanwhile, the zero-point of scalar curvature is uniquely determined by $b(V, Q)$ term, which means that it will not be changed by GUP correction.

However, we should also notice that due to GUP correction, $v_{min}=2\alpha$, so if
\begin{equation}
\label{eq:38}
\frac{\alpha}{Q} \geqslant \sqrt{2},
\end{equation}
there is no sign-changing point, indicating only attractive interaction exists in the black hole microstructure. The GUP correction therefore gives a new limitation for repulsive interaction existing in the black hole microstructure, where the value of $\alpha^{2}/Q^{2}$ will determine whether there is an actual sign-changing point.

On the other hand, although $R$ is positive in some regions when $\alpha^{2}/Q^{2} < 2$, we should examine whether these regions are thermodynamically stable. In the followings, we separately check this in the VdW-like PT case and RPT case. 

\subsection{Van Der Waals-Like Phase Transition Case}
\begin{figure}[h]
\centering
\subfloat[$\alpha^{2}/Q^{2}= 0$]
{\begin{minipage}[b]{.33\linewidth}
\centering
\includegraphics[height=5cm]{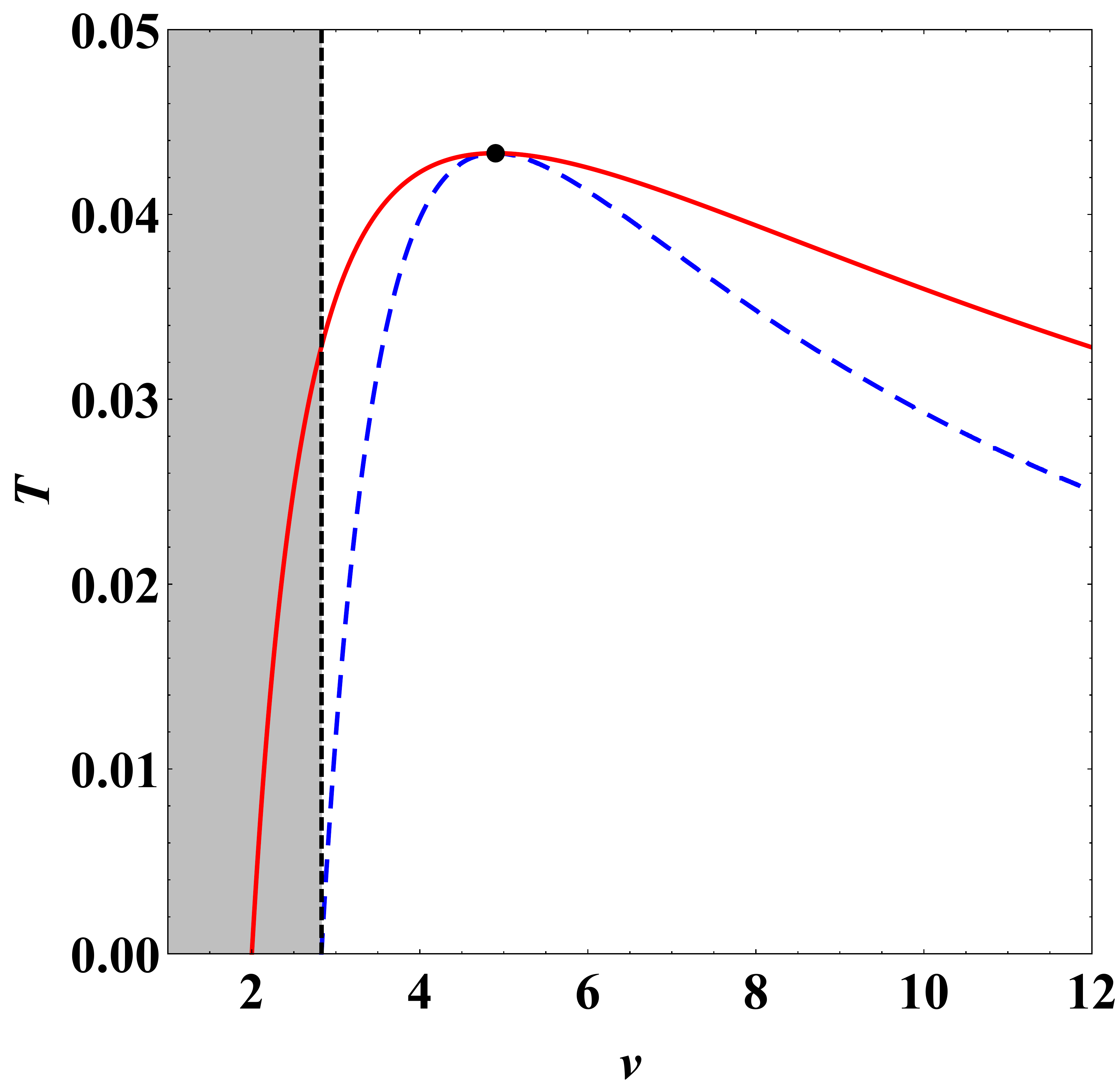}
\label{fig:Fig.9(a)}
\end{minipage}
} 
\subfloat[$\alpha^{2}/Q^{2}=0.95$]
{\begin{minipage}[b]{.33\linewidth}
\centering
\includegraphics[height=5cm]{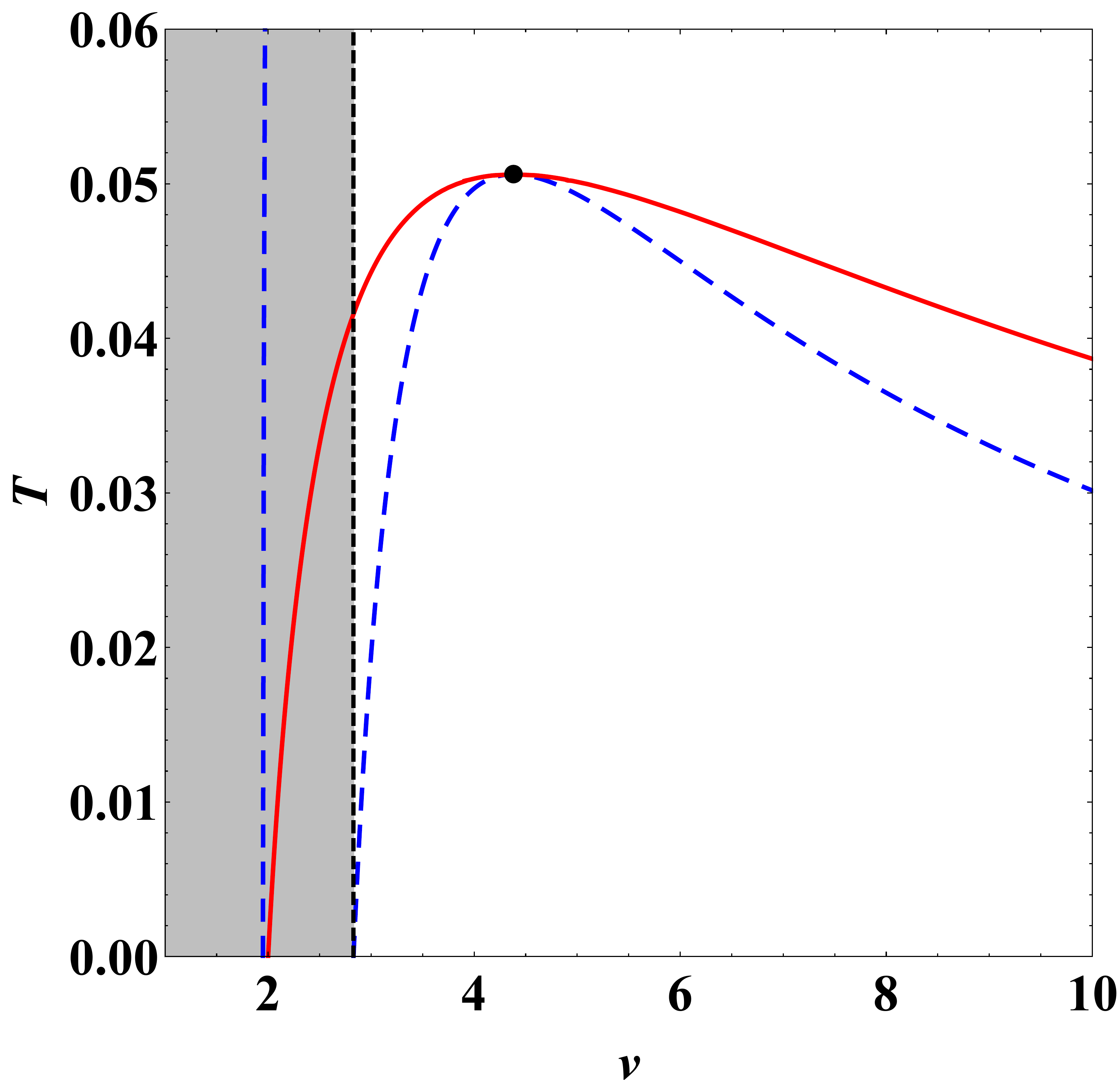}
\label{fig:Fig.9(b)}
\end{minipage}
}
\subfloat[$\alpha^{2}/Q^{2}=1$]
{\begin{minipage}[b]{.33\linewidth}
\centering
\includegraphics[height=5cm]{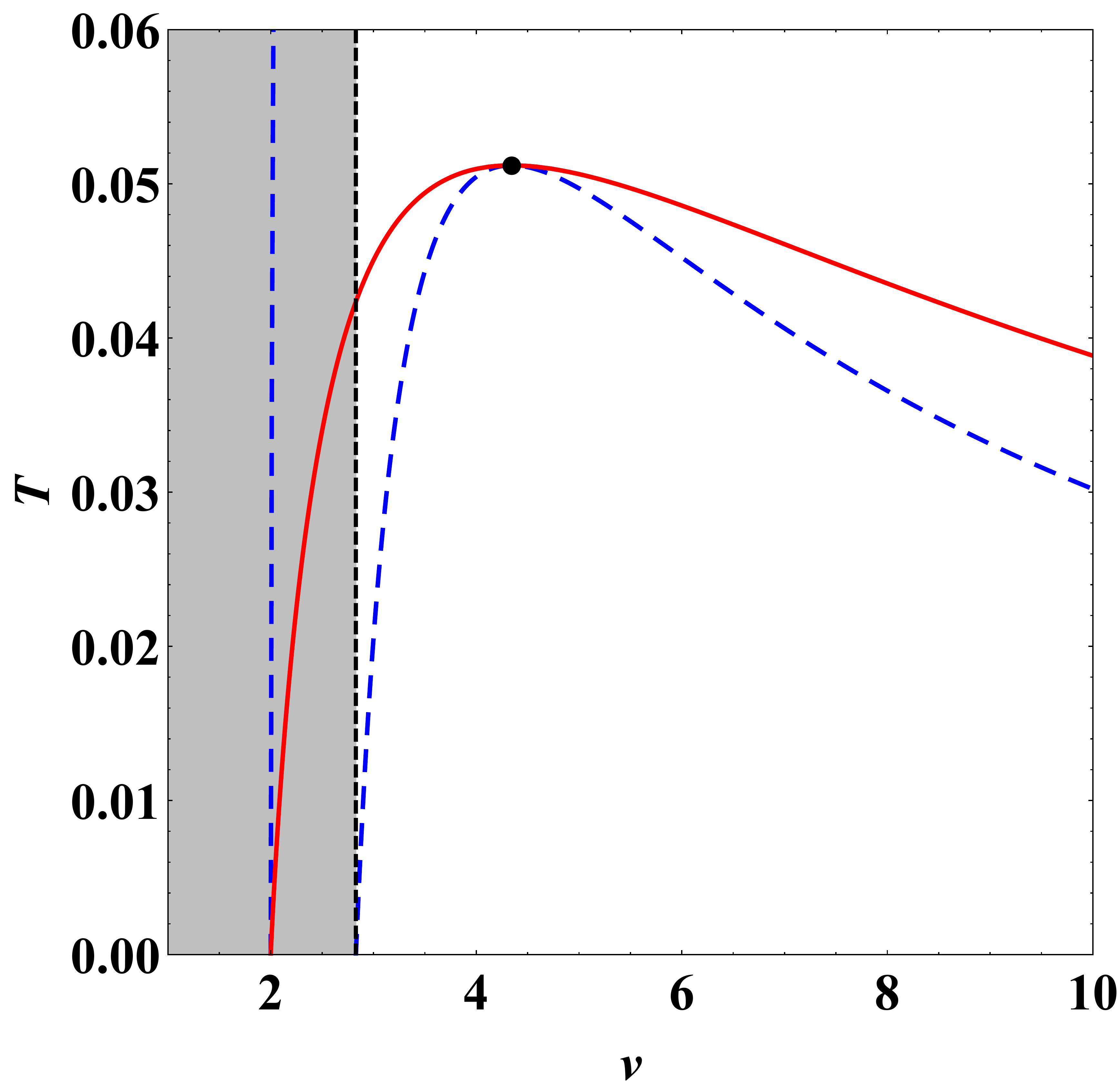}
\label{fig:Fig.9(c)}
\end{minipage}
}
\quad
\subfloat[$\alpha^{2}/Q^{2}= 0$]
{\begin{minipage}[b]{.33\linewidth}
\centering
\includegraphics[height=5cm]{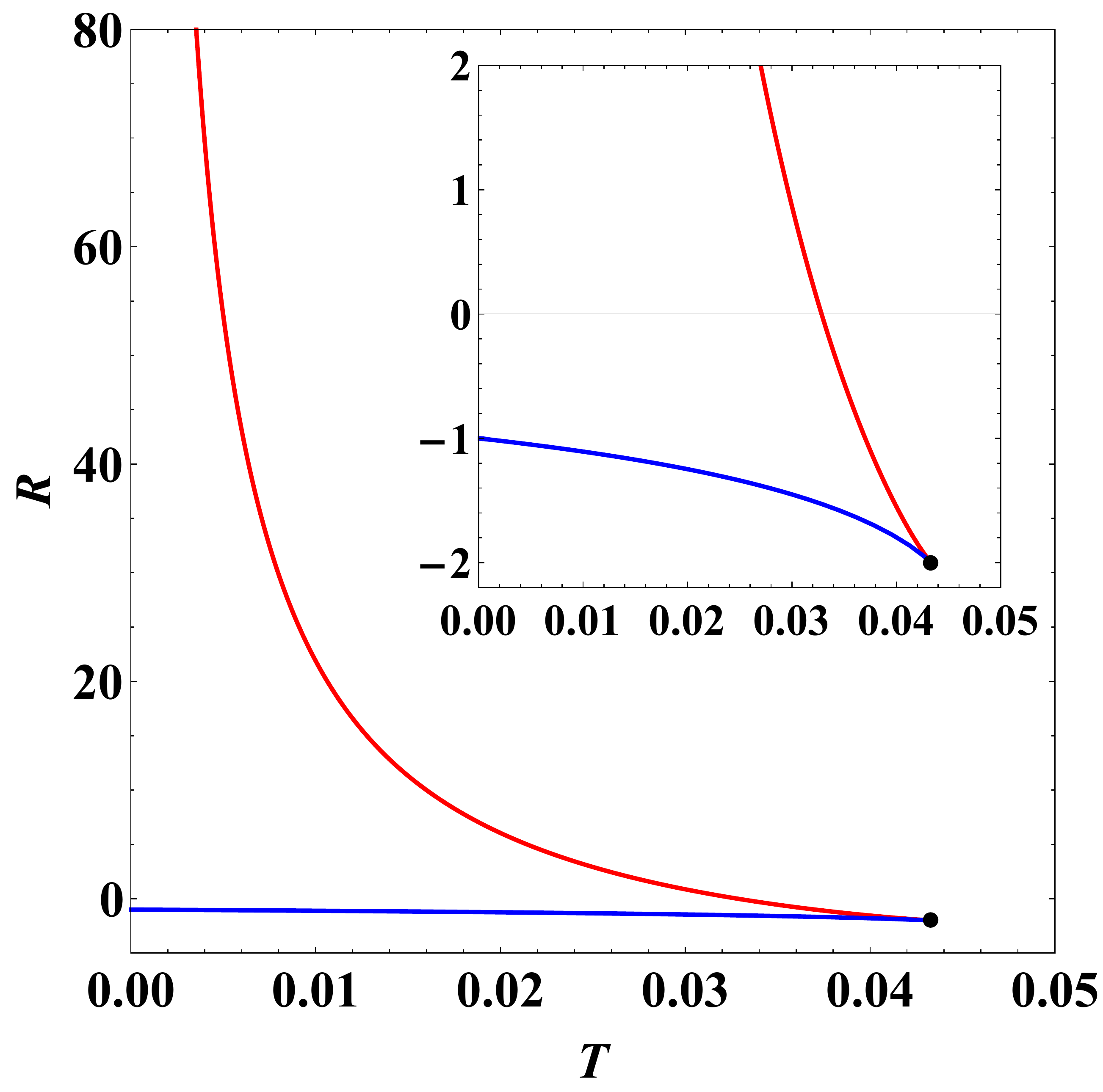}
\label{fig:Fig.9(d)}
\end{minipage}
} 
\subfloat[$\alpha^{2}/Q^{2}=0.95$]
{\begin{minipage}[b]{.33\linewidth}
\centering
\includegraphics[height=5cm]{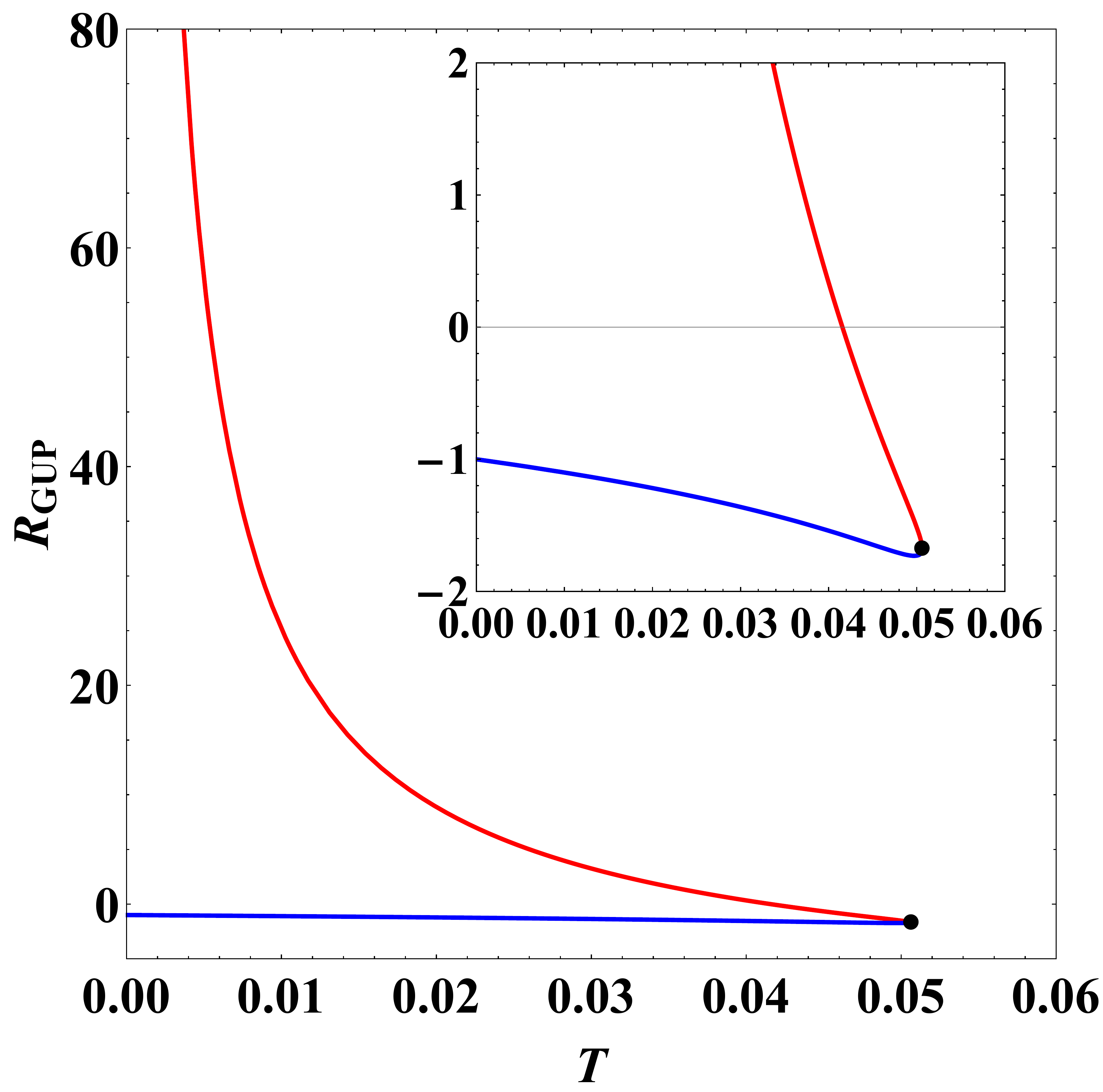}
\label{fig:Fig.9(e)}
\end{minipage}
}
\subfloat[$\alpha^{2}/Q^{2}=1$]
{\begin{minipage}[b]{.33\linewidth}
\centering
\includegraphics[height=5cm]{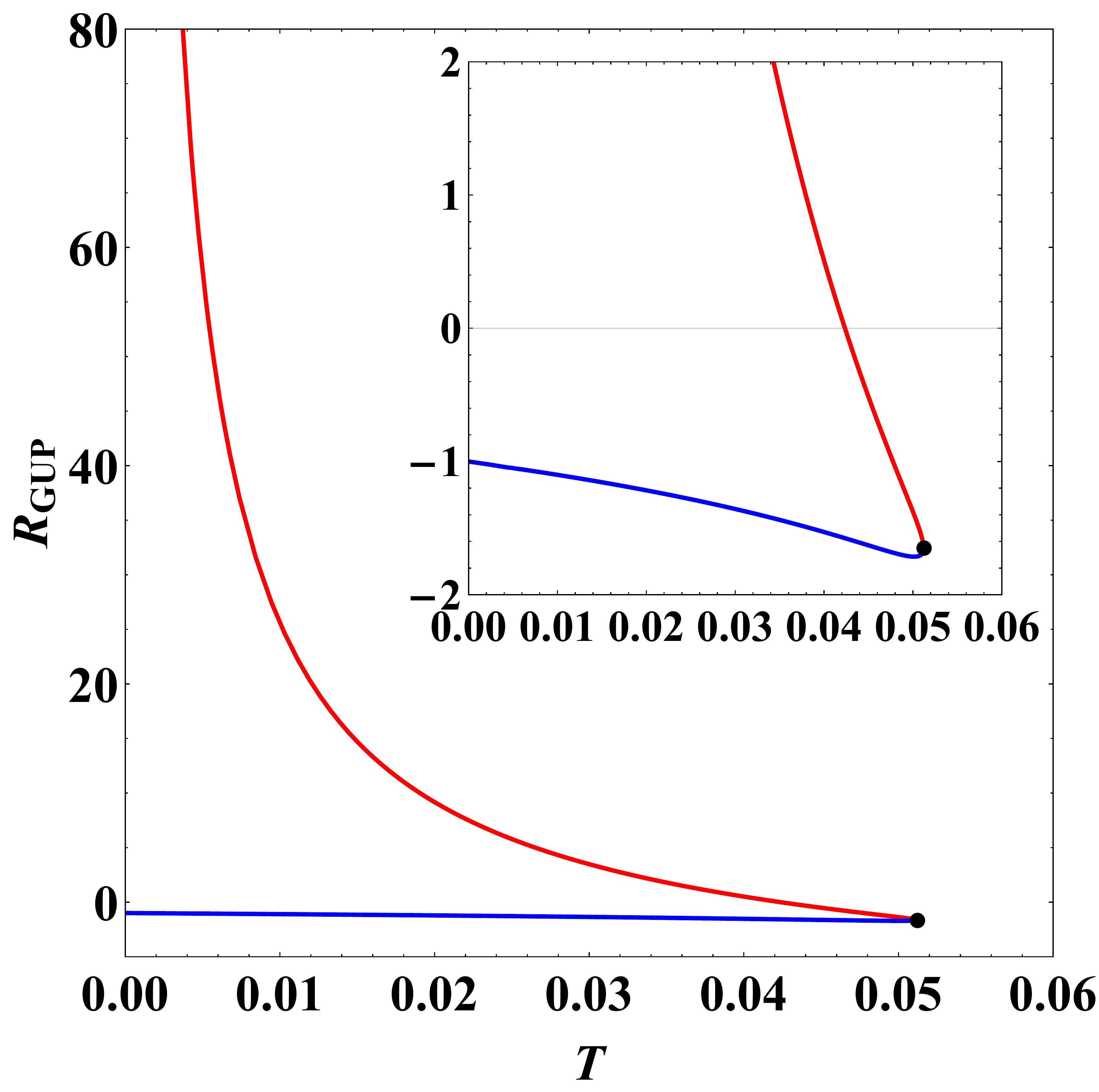}
\label{fig:Fig.9(f)}
\end{minipage}
}
\caption{\label{fig:Fig.9}(a-c) The phase transition curves (red solid curves), spinodal curves (blue dashed lines), and sign-changing curves (black dashed lines) for the VdW-like PT case with different values of $\alpha^{2}/Q^{2}$. The positive $R$ regions, i.e. the regions of $v<2\sqrt{2}Q$, have been shadowed. (d-f) The behavior of scalar curvature along the coexistence curves. The red and blue solid lines represent the IBH (SBH) and LBH branches respectively. Insets describe the sign-changing behavior of the SBH branches. We have set $Q=1$.}
\end{figure}

We depict the phase transition curves (red solid curves), spinodal curves (blue dashed lines), and sign-changing curves (black dashed lines) in Figs. \ref{fig:Fig.9}(\subref*{fig:Fig.9(a)}--\subref*{fig:Fig.9(c)}). The positive $R$ regions, i.e., the regions of $v<2\sqrt{2}Q$, have been shadowed. In the case where $\alpha^{2}/Q^{2}\leqslant 1$, there is a positive $R$ region for the IBH (or SBH for $\alpha^{2}/Q^{2}= 0$), which means repulsive interaction may exist in the black hole.

Now we examine the behavior of scalar curvature along the coexistence curves. With the normalized scalar curvature shown in Eq. (\ref{eq:28}), the critical behavior in the charged AdS black hole has been discussed in Refs. \cite{Wei:2019uqg,Wei:2019yvs}. They showed that along the coexistence saturated LBH curve, $R_{N}$ is always negative, whereas along the coexistence saturated SBH $R_{N}$ is positive for low temperature and negative for high temperature. On the other hand, at the critical point, both $R_{N}$ of SBH and LBH branches diverge to negative infinity. More interestingly, 
they found that the critical behaviors for a charged AdS black hole and the VdW fluid are identical. 

Additionally, with another version of the scalar curvature Eq. (\ref{eq:32}), similar behavior can also be found. Fig. \ref{fig:Fig.9}(\subref*{fig:Fig.9(d)}) shows clearly that, in the LBH phase, the dominant interaction is always attractive ($R<0$), while in the SBH phase, it is repulsive ($R>0$) at low temperatures and then becomes attractive ($R<0$) at high temperatures. However, the divergence occurs at $T\rightarrow 0$ instead of the critical point. We shall not discuss the difference in the divergence here, but focus on the sign of scalar curvature which indicates the interaction type of black hole microstructure. At least on this point, the two different scalar curvatures give the same result, as discussed before. Interestingly, for the LBH branch, $R \rightarrow -1$ instead of 0 when $T\rightarrow 0$ (i.e. $r_{h} \rightarrow +\infty$, see the coexistence curve shown in Fig. \ref{fig:Fig.9}(\subref*{fig:Fig.9(a)})), which is comparable to the scalar curvature at critical piont ($R = -2$). With the analytic expression for the coexistence curve of the black hole, we can find this feature is independent of the value of $Q$. In fact, using the normalized scalar curvature Eq. (\ref{eq:28}), one can also find $R_{N} \rightarrow -3/2$ instead of 0 for the LBH branch when $T\rightarrow 0$, whose magnitude is comparable to the scalar curvature for the SBH branch at the same temperature ($R_{N}\rightarrow1/2$). This phenomenon suggests that even at infinite thermodynamic volume, non-negligible attractive interactions may arise between the black hole microscopic ingredients, which differs from a VdW fluid system \cite{Wei:2019yvs}.

For $0<\alpha^{2}/Q^{2}\leqslant 1$, the behavior of scalar curvature along the phase transition curve is similar to the $\alpha^{2}/Q^{2}=0$ case, as shown in Figs. \ref{fig:Fig.9}(\subref*{fig:Fig.9(e)}) and \ref{fig:Fig.9}(\subref*{fig:Fig.9(f)}). In summary, in the case of VdW-like phase transition, the LBH phase is always dominated by attractive interaction, while the IBH phase contains both attractive and repulsive interactions. It is worth noting that from numerical results, the scalar curvature also takes $-1$ instead of 0 for the coexistence saturated LBH when $T\rightarrow 0$, which is not influenced by the value of $\alpha^{2}/Q^{2}$.

\subsection{Reentrant Phase Transition Case}
\begin{figure}
\centering
\subfloat[$\alpha^{2}/Q^{2}=1.4$]
{\begin{minipage}[b]{.33\linewidth}
\centering
\includegraphics[height=4.9cm]{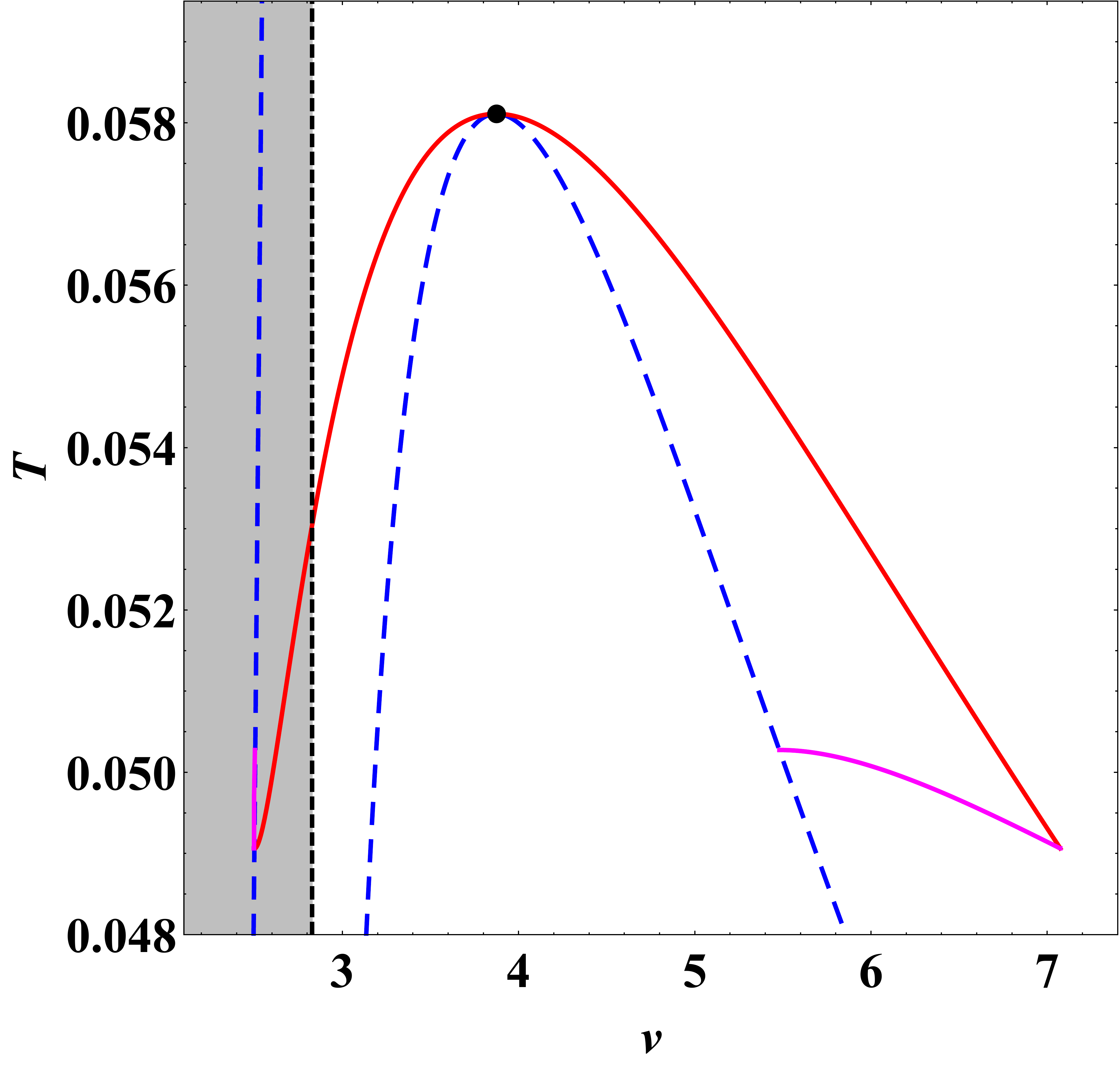}
\label{fig:Fig.10(a)}
\end{minipage}
} 
\subfloat[$\alpha^{2}/Q^{2}=1.5$]
{\begin{minipage}[b]{.33\linewidth}
\centering
\includegraphics[height=4.9cm]{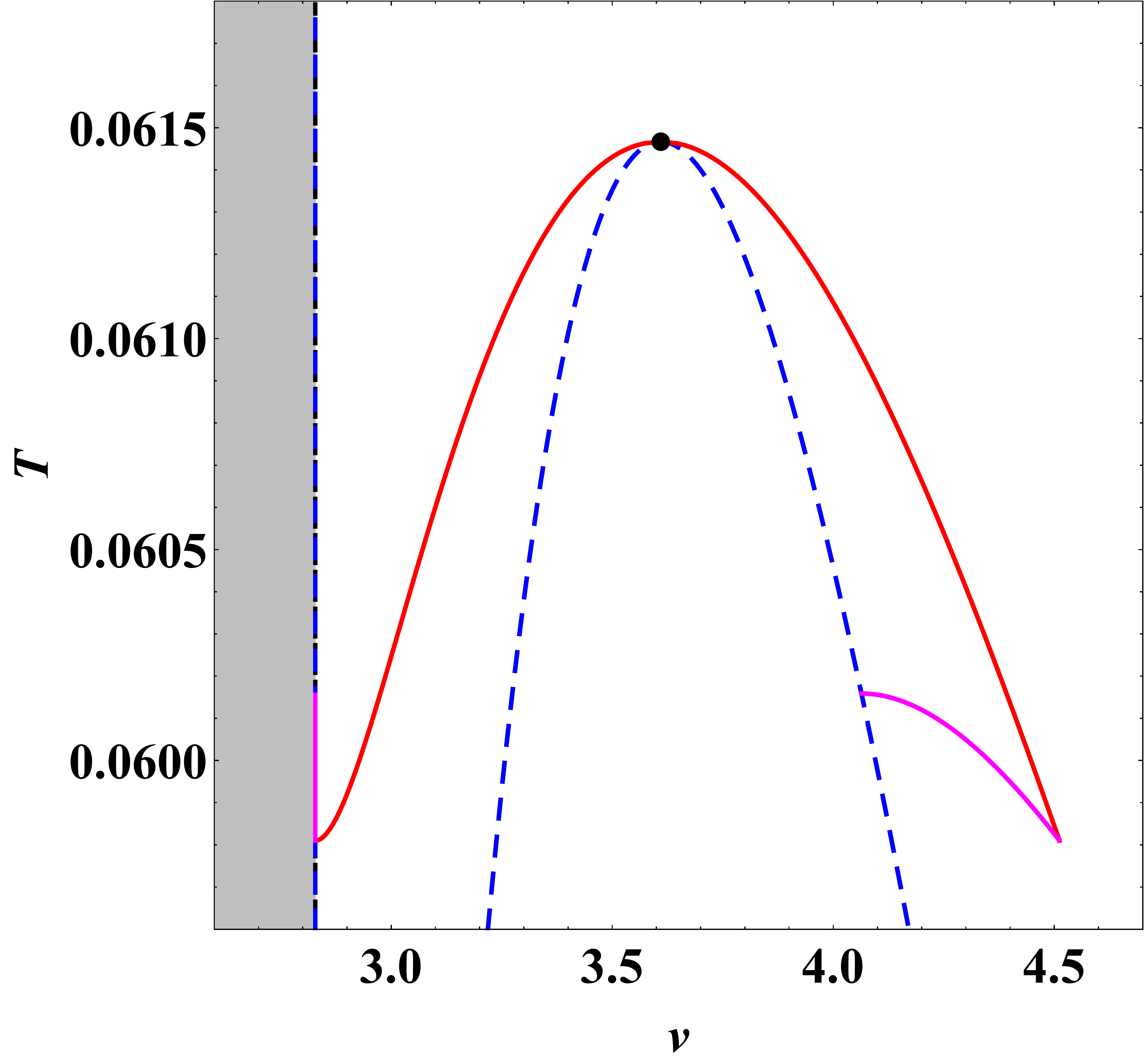}
\label{fig:Fig.10(b)}
\end{minipage}
}
\subfloat[$\alpha^{2}/Q^{2}=1.51$]
{\begin{minipage}[b]{.33\linewidth}
\centering
\includegraphics[height=4.9cm]{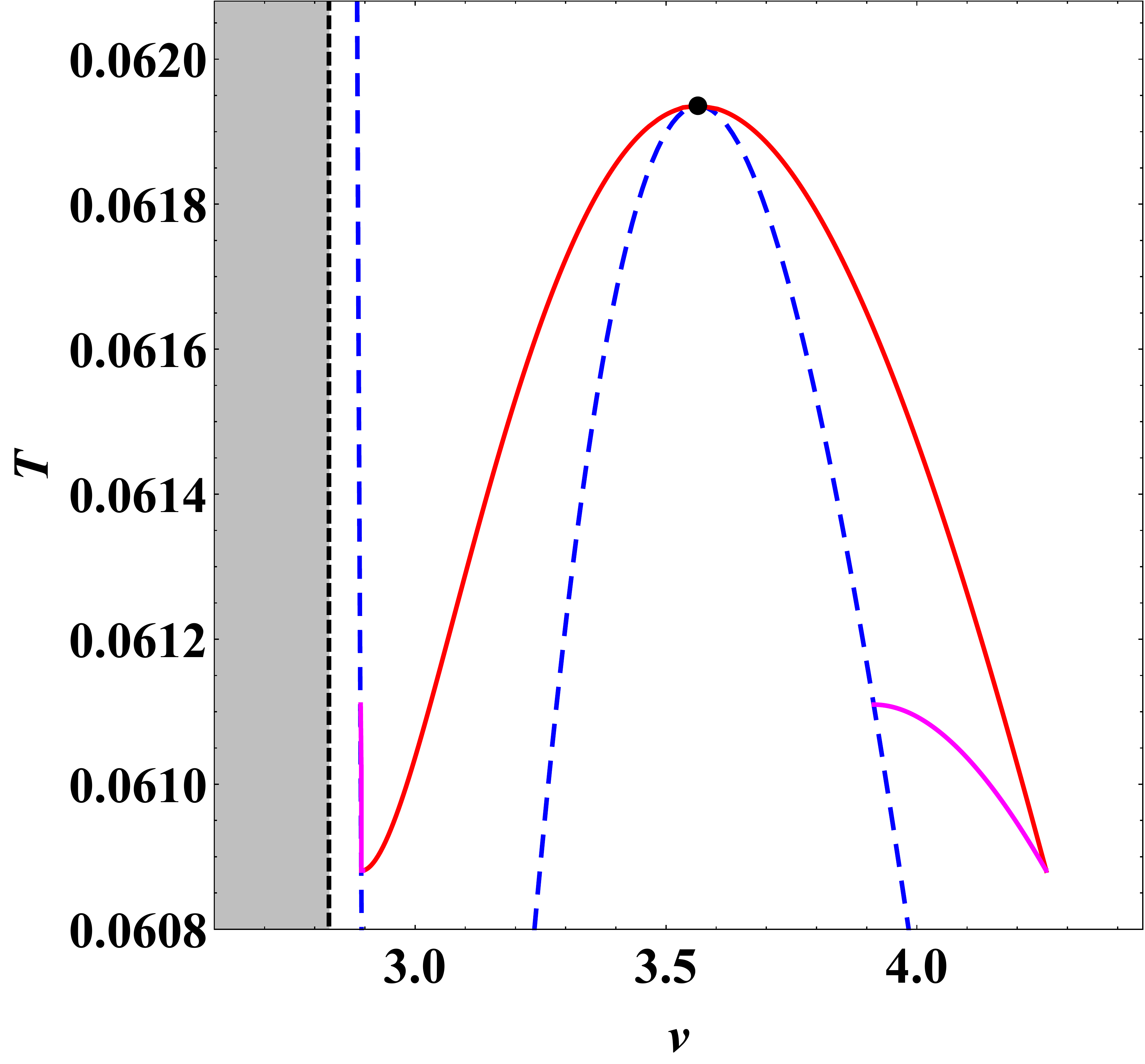}
\label{fig:Fig.10(c)}
\end{minipage}
}
\quad
\subfloat[$\alpha^{2}/Q^{2}=1.4$]
{\begin{minipage}[b]{.33\linewidth}
\centering
\includegraphics[height=4.9cm]{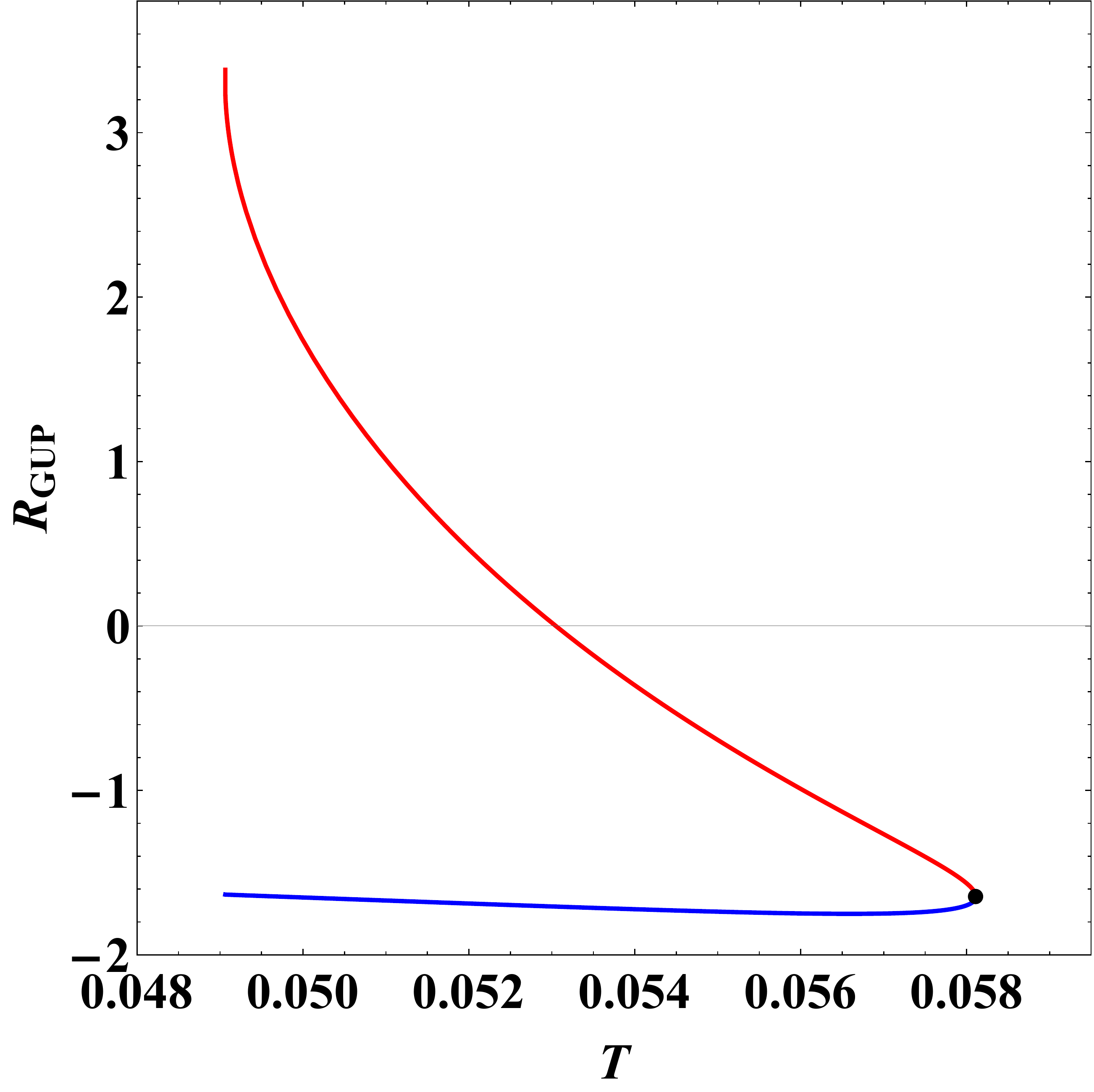}
\label{fig:Fig.10(d)}
\end{minipage}
} 
\subfloat[$\alpha^{2}/Q^{2}=1.5$]
{\begin{minipage}[b]{.33\linewidth}
\centering
\includegraphics[height=4.9cm]{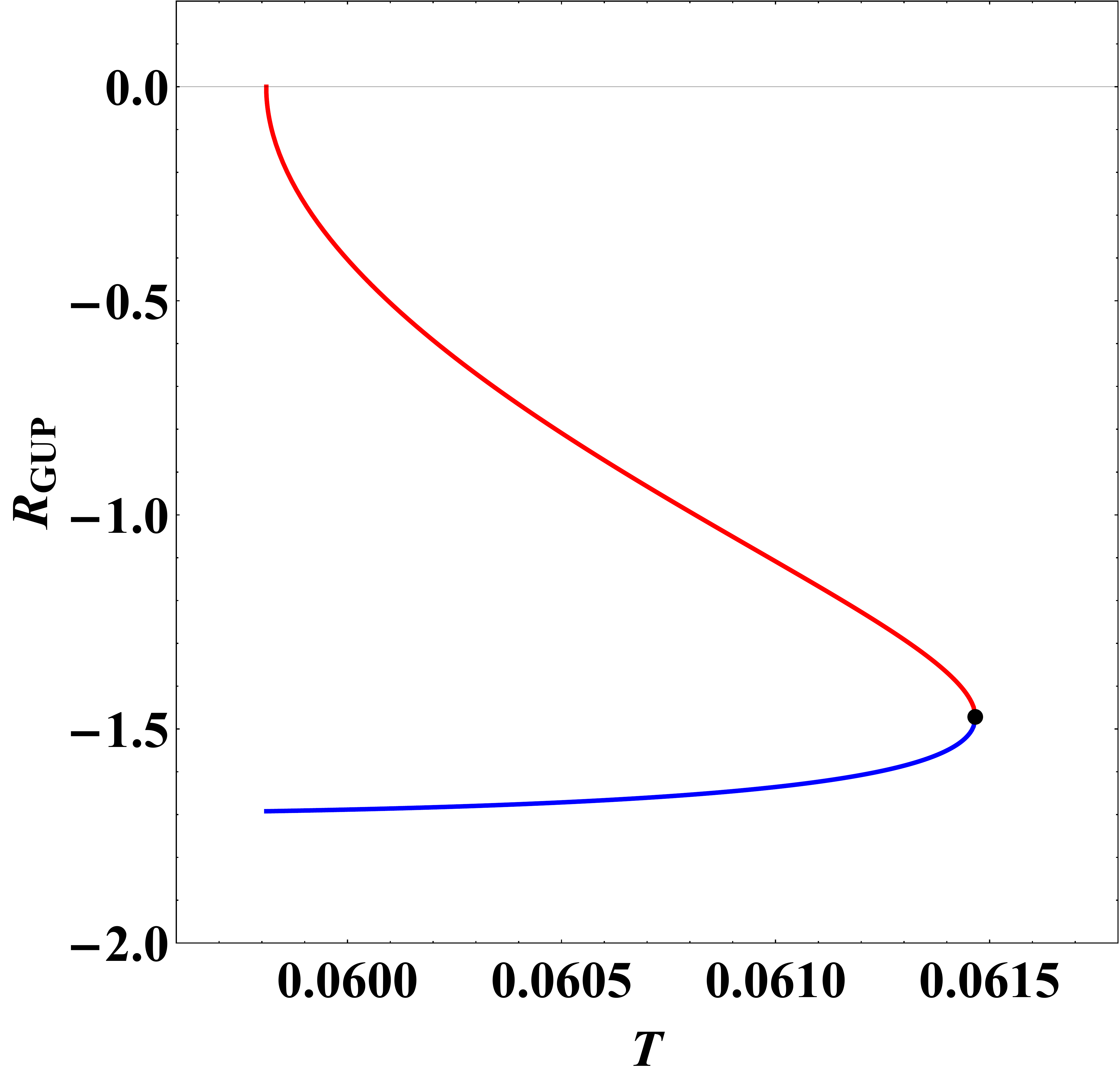}
\label{fig:Fig.10(e)}
\end{minipage}
}
\subfloat[$\alpha^{2}/Q^{2}=1.51$]
{\begin{minipage}[b]{.33\linewidth}
\centering
\includegraphics[height=4.9cm]{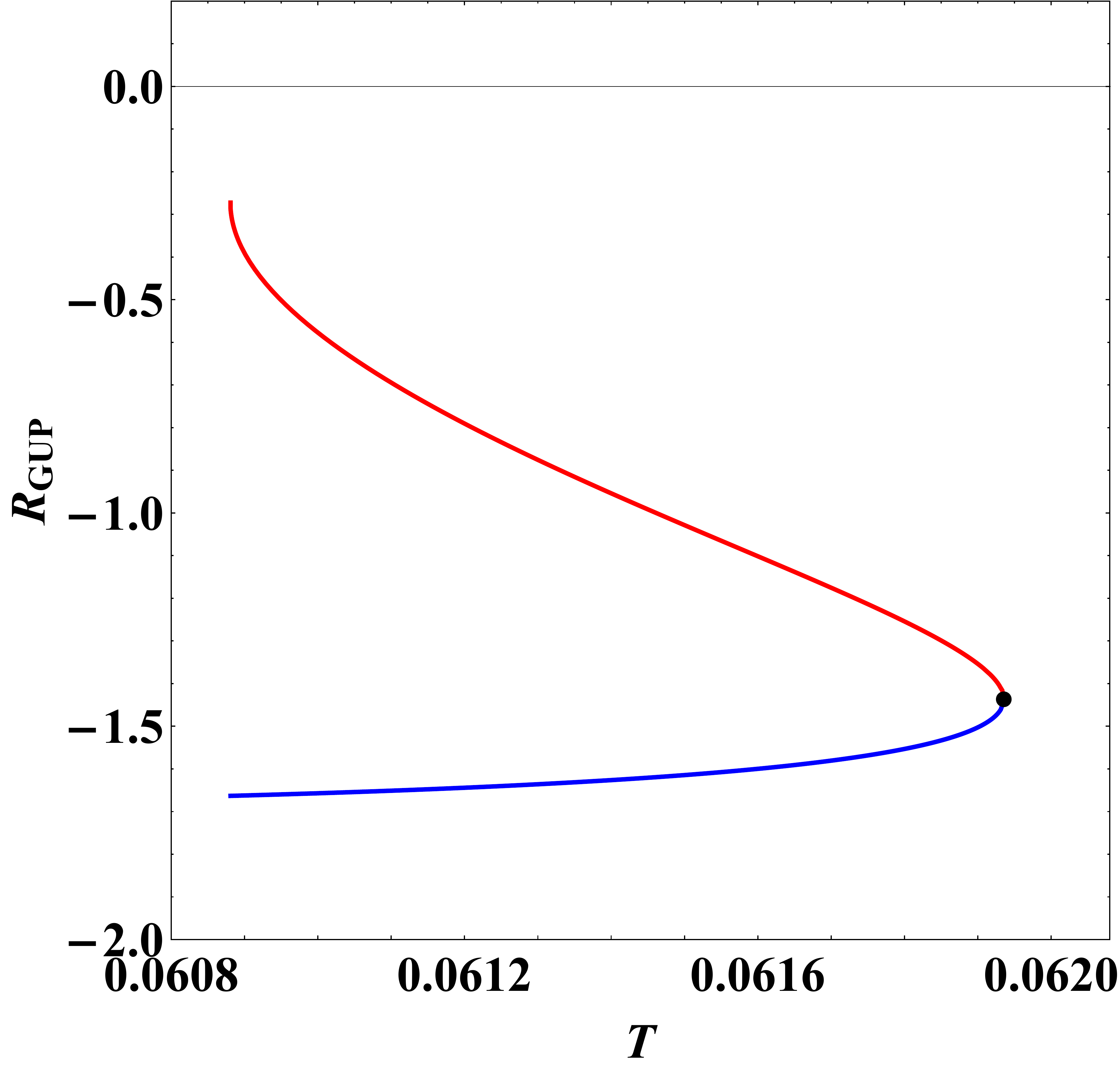}
\label{fig:Fig.10(f)}
\end{minipage}
}
\caption{\label{fig:Fig.10}(a-c) The zeroth-order (magenta solid curves) and first-order (red solid curves) phase transition curves, spinodal curves (blue dashed lines), and sign-changing curves (black dashed lines) for the RPT case with different values of $\alpha^{2}/Q^{2}$. The positive $R$ regions, i.e., the regions of $v<2\sqrt{2}Q$, have been shadowed. (d-f) The behavior of scalar curvature along the coexistence curves (first-order phase transition curves). The red and blue solid lines represent the IBH and LBH branches, respectively. We have set $Q=1$.}
\end{figure}

We display the zeroth-order (magenta solid curves) and first-order (red solid curves) phase transition curves, spinodal curves (blue dashed lines), and sign-changing curves (black dashed lines) in Figs. \ref{fig:Fig.10}(\subref*{fig:Fig.10(a)}--\subref*{fig:Fig.10(c)}). For $\alpha^{2}/Q^{2}>1.5$, the positive $R$ regions do not coincide with any stable phase. As mentioned in Sec. \ref{section3}, the stable phases are on the right side of line $v=4\sqrt{3} \alpha /3$ when $\alpha^{2}/Q^{2}>1.5$. For these parameter regions, the sign-changing line $r_{h}=\sqrt{2}Q$ is on the left side of line $v=4\sqrt{3} \alpha /3$, and thus there are no stable phases with repulsive interaction.

In fact, $\alpha^{2}/Q^{2}=1.5$ is the critical value for existing repulsive interaction in stable phases. The sign-changing line $v=2\sqrt{2}Q$ overlaps with the line $v=4\sqrt{3} \alpha /3$, as shown in Fig. \ref{fig:Fig.10}(\subref*{fig:Fig.10(b)}). When $\alpha^{2}/Q^{2}<1.5$, the line $v=4\sqrt{3} \alpha /3$ moves to the left side of the sign-changing line $v=2\sqrt{2}Q$, and there will be stable phases with positive scalar curvature as in Fig. \ref{fig:Fig.10}(\subref*{fig:Fig.10(a)}).

The behavior of scalar curvature along the coexistence curves (i.e. first-order phase transition curves) is shown in Figs. \ref{fig:Fig.10}(\subref*{fig:Fig.10(d)}--\subref*{fig:Fig.10(f)}). In all cases, the LBH phase always has dominant attractive interaction. For $\alpha^{2}/Q^{2}<1.5$, similar to the VdW-like PT case, a transition between attractive and repulsive interactions appears in the IBH phase. $\alpha^{2}/Q^{2}=1.5$ is a turning point for repulsive interaction existing in the IBH phase, where the scalar curvature of IBH phase decays from 0 to a negative value with the increasing of temperature. While $\alpha^{2}/Q^{2}>1.5$, the scalar curvature of IBH phase is always negative, implying the dominance of attractive interaction.

It has been argued that the interaction type in IBHs (or SBHs) near the first-order phase transition line, differs for VdW-like PT and RPT. In the VdW-like case, a transition between attractive and repulsive interactions can be found \cite{Wei:2019yvs,NaveenaKumara:2020jmk,Wei:2020poh,Wei:2021lmo,Gogoi:2021syo} whereas for the RPT case the dominant interaction is always attractive \cite{NaveenaKumara:2020biu,KordZangeneh:2017lgs} or repulsive \cite{Chen:2018icg}. In other words, in the RPT case, the IBHs (or SBHs) behave like bosonic gas or fermionic gas, while in the VdW-like PT case, they behave like quantum anyon gas. However, for the GUP-corrected charged AdS black hole, the IBH branch gradually shifts to the negative scalar curvature region with the increase in $\alpha^{2}/Q^{2}$ (see Fig. \ref{fig:Fig.10}), resulting in the quantum anyon gas like IBHs for $\alpha^{2}/Q^{2} <1.5$ and the Bosonic gas like IBHs for $\alpha^{2}/Q^{2} >1.5$, although they are both in the RPT case. In Ref. \cite{Ye:2022uuj}, the quantum anyon gas like SBHs in the RPT case are also found.

\section{Conclusions and discussions}\label{section5}
The microstructure of the GUP-corrected charged AdS black hole has been investigated considering the Ruppeiner geometry and phase transition behavior. It was shown that the microstructure of the black hole depends on the ratio between GUP parameter $\alpha$ and electric charge $Q$.

For a small ratio $\alpha^{2}/Q^{2} \leqslant 1$, the typical VdW-like phase transition is present in the system. The comparison shown in Fig. \ref{fig:Fig.6} indicates that the black holes with and without GUP correction have similar phase structures. However, the SBH phase is replaced by the IBH phase for there is another unstable branch in the small black hole region. Meanwhile, the region of no black hole arises as a reuslt of the GUP correction. The study of Ruppeiner scalar curvature shows that in addition to the dominant attractive interaction, a repulsive interaction between microscopic ingredients of IBHs exists in a small parameter range $r<\sqrt{2}Q$ (see Fig. \ref{fig:Fig.9}). As might be expected for the LBHs, attractive interaction is dominant as their microscopic ingredients are supposed to be far from each other \cite{KordZangeneh:2017lgs}. Along the coexistence curve, the LBH phase always exhibits dominant attractive interaction, while a transition between the dominant repulsive and attractive interactions is found in the SBH phase. This result indicates that the microscopic interaction of the black hole undergoes drastic change during the phase transition.

Further increasing the ratio between $\alpha$ and $Q$ to $1<\alpha^{2}/Q^{2}<1.535$, three cases of RPT are displayed by the black hole, namely $1<\alpha^{2}/Q^{2}<1.5$, $\alpha^{2}/Q^{2}=1.5$ and $1.5<\alpha^{2}/Q^{2}<1.535$. Their phase structures are considerably similar, which can be directly seen from Fig. \ref{fig:Fig.8}. However, the Ruppeiner geometry study shows that their microstructures are different, as demonstrated in Fig. \ref{fig:Fig.10}. Specifically, for $1<\alpha^{2}/Q^{2}<1.5$, repulsive interaction exists in the IBH phase, while in other parameter spaces the IBH only has dominant attractive interaction. For the LBH phase, the attractive interaction is always dominated in all cases. Along the coexistence curve, a transition between dominant attractive and repulsive interactions for the IBH branch is found when $1<\alpha^{2}/Q^{2}<1.5$, accompanied by the dominant attractive interaction in the LBH branch. In other cases, however, attractive interaction is always dominated, both in IBH and LBH branches. This result is quite different from the RPT case shown by several other black holes \cite{NaveenaKumara:2020biu,KordZangeneh:2017lgs,Chen:2018icg}, where there are only attractive or repulsive interaction in black holes.

For a large ratio $\alpha^{2}/Q^{2}>1.535$, the LBH phase is globally stable, hence there is no phase transition in the system. As discussed in the RPT case, $\alpha^{2}/Q^{2}=1.5$ is the critical value for existing repulsive interaction in stable phases. Therefore, for the case of no phase transition, only attractive interaction exists among microscopic ingredients of stable black holes. These features reveal that the system behaves like a Schwarzschild-AdS black hole \cite{Xu:2020gud}.

To conclude, the increasing ratio of GUP parameter $\alpha$ and electric charge $Q$ will reduce the presented repulsive interaction in charged AdS black holes. This result can be qualitatively understood from the perspective of black hole molecules \cite{Wei:2015iwa}. The GUP effect results in a minimal specific volume $v_{min}=2 \alpha$, which may be identified as the volume of black hole molecules. On the other hand, the GUP effect does not change the second and third terms of the state equation Eq. (\ref{eq:15}), suggesting the interaction form of black hole molecules is maintained. From the Ruppeiner geometry analysis in Sec. \ref{section4}, the turning point between the dominant repulsive and attractive interactions is uniquely determined 
as $v=2 \sqrt{2} Q$. Thus, for a larger ratio between $\alpha$ and $Q$, the repulsive interaction tends to decrease and the attractive interaction will be dominant.

Further investigations of AdS black holes with the minimal length effect may reveal more interesting microstructures. In addition, we hope these investigations will help us understand the nature of a black hole, from the perspective of black hole thermodynamics and the underlying black hole microstructures.

\begin{acknowledgments}
We are grateful to Peng Wang,  Feiyu Yao and Chen Su for useful discussions and valuable comments. This work is supported by NSFC (Grant No.11947408 and 12047573).
\end{acknowledgments}

\end{document}